\documentclass[12pt]{article}

\usepackage{epsfig}
\usepackage{amssymb}
\usepackage{amsmath}
\usepackage{amsfonts}
\usepackage{url}
\usepackage{hyperref}

\def\vec#1{{\bf #1}}

\newcommand{\be}{\begin{equation}}
\newcommand{\ee}{\end{equation}}
\newcommand{\beq}{\begin{equation}}
\newcommand{\eeq}{\end{equation}}
\newcommand{\bea}{\begin{eqnarray}}
\newcommand{\eea}{\end{eqnarray}}

\newcommand{\lp}{\left(}
\newcommand{\rp}{\right)}
\newcommand{\as}{\alpha_s}

\def\frac#1#2{{{#1}\over {#2}}}
\def\gsim{\mathrel{\rlap{\lower4pt\hbox{\hskip1pt$\sim$}}
    \raise1pt\hbox{$>$}}}         
\def\lsim{\mathrel{\rlap{\lower4pt\hbox{\hskip1pt$\sim$}}
    \raise1pt\hbox{$<$}}}         

\newcommand{\draft}[1]{}

\def\Aslash{\not{\hbox{\kern-4pt $A$}}}
\def\Eslash{\not{\hbox{\kern-4pt $E$}}}

\newcommand{\bi}{\begin{itemize}}
\newcommand{\ei}{\end{itemize}}
\newcommand{\ben}{\begin{enumerate}}
\newcommand{\een}{\end{enumerate}}
\def\href#1{(\ref{#1})}
\def \as {\alpha_s}
\def\gsim{\mathrel{\rlap{\lower4pt\hbox{\hskip1pt$\sim$}}
    \raise1pt\hbox{$>$}}}         
\def\lsim{\mathrel{\rlap{\lower4pt\hbox{\hskip1pt$\sim$}}
    \raise1pt\hbox{$<$}}}         
\def \asb {\bar \alpha_s}

\def \zbar {\bar z}
\def \d {{d }}
\def \msb {\overline{\textsc{MS}}}

\begin{document}
\begin{flushright}
IFUM-966-FT\\
MAN/HEP/2010/13
\end{flushright}

\begin{center}
{\Large \bf
Small $x$  resummation of rapidity distributions:\\
the case of Higgs production}
\vspace*{1.5cm}

Fabrizio Caola$^{a}$, Stefano Forte$^{a}$ and Simone Marzani$^b$
\\
\vspace{0.3cm}  {\it
{}$^a$Dipartimento di Fisica, Universit\`a di Milano and
INFN, Sezione di Milano,\\
Via Celoria 16, I-20133 Milano, Italy\\ \medskip
{}$^b$School of Physics \& Astronomy, University of Manchester,\\
Oxford Road, Manchester, M13 9PL, England, U.K.}\\
\vspace*{1.5cm}

\bigskip
\bigskip

{\bf \large Abstract:}
\end{center}
We provide a method for the all order computation of 
small $x$ contributions at the leading logarithmic level
to cross-sections which are differential in
rapidity. The method is based on a
generalization to rapidity distributions of the high energy (or $k_T$)
factorization theorem hitherto proven for inclusive 
cross-sections. We apply the method to Higgs production in gluon-gluon
fusion, both with finite top mass and in the infinite mass limit:
in both cases, we determine all-order resummed
expressions, as well as explicit expressions for the leading small $x$
terms up to NNLO. We use our result to construct an explicit approximate
analytic expression of the finite--mass NLO rapidity distribution 
and an estimate of finite--mass corrections at NNLO.

\vfill
\begin{flushleft}
October 2010
\end{flushleft}

\clearpage

\tableofcontents
\clearpage

\section{Introduction}
Techniques for the computation of small $x$ logarithmically enhanced
contributions to hard QCD cross-sections to all orders have been
available for some time, and have been successfully applied to a
variety of hard processes: heavy quark photo- and
lepto-production~\cite{CataniHQ}, Deep-Inelastic
Scattering~\cite{CataniDIS}, and more recently hadroproduction
processes, including heavy quarks~\cite{BallEllis},
Higgs (both in the infinite top mass limit~\cite{HautmannHiggs}
and for finite $m_{\rm top}$~ \cite{SimoneHiggs,SimoneHiggsProc}),  
Drell-Yan~\cite{SimoneDY} and prompt-photon~\cite{GiovanniPhoton}.
In each of these cases, the leading contribution to the hard cross-section
in the  $x\to0$ limit has been determined to all orders in the strong
coupling $\alpha_s(Q^2)$, where $Q^2$ is the hard scale of the
process (typically related to the invariant mass of the final state
for hadroproduction processes), and  $x$ is the dimensionless ratio of
this scale to the available (partonic) center-of-mass energy  $x= Q^2/s$.

The interest in these results is twofold. First, they provide
information  on  unknown higher--order contributions (or cross-checks
on known ones): for example, the NNLO leading small $x$
contribution to Higgs production in the finite $m_{\rm top}$
case~\cite{SimoneHiggs} was used in  Ref.~\cite{SimoneHarlander}  to
construct an approximate expression for the yet unknown
full NNLO term by combining it with an expansion 
in $1/m_{\rm top}$~\cite{harlander,pak}. Second, they can be used to construct 
all order resummed results, which may be necessary in the kinematic
region where $\as \ln\frac{1}{x}\sim 1$ so that the conventional
perturbative expansion becomes unstable. For many processes mentioned
above, such as Deep-Inelastic Scattering or Drell-Yan,
the Born cross-section is a quark-induced $O(\alpha_s^0)$
electroweak process (parton model), so
that the leading-log resummation of hard cross-sections
is only consistent (i.e. resummed results are
only factorization--scheme invariant) if combined with a
next-to-leading log resummation of evolution equations. Such a
consistent resummation is possible because not only
leading~\cite{BFKL} but also next-to-leading~\cite{FL}
$\ln\frac{1}{x}$ contributions to evolution equations are known since
some time, as well as methods to use resummed evolution equations in a
way compatible with physical constraints~\cite{CCSS,ABF}, and to
combine solutions to resummed evolution equations with resummed hard
partonic cross-sections~\cite{Ball:2007ra,Altarelli:2008aj}.

The usefulness of these results for phenomenology has been hampered so
far by the fact that they, and the computational techniques on which
they are based, are only available for fully inclusive
cross-sections. This is likely to become a very serious limitation at
the LHC where, for instance, the LHCb experiment will measure Drell-Yan
rapidity distributions at very small invariant masses (i.e. at very 
small $x$) but in a limited rapidity
range~\cite{McNulty:2009zz}. It is the purpose of this paper to
overcome this limitation by extending the resummation
formalism to rapidity distributions.

Available calculations at small $x$ are based on a suitable
formulation of QCD factorization which is consistent with all-order leading
log~$x$ resummation, called high energy or
$k_T$~factorization~\cite{CataniHQ,Catani:1993ww}. Generalization of
this factorization to the less inclusive case is a prerequisite for
resummation, and it will be accomplished here. We will proceed as
follows. First, in Sect.~\ref{inclusive} we will rederive the standard
high energy factorization for inclusive cross-sections
of Refs.~\cite{CataniHQ,Catani:1993ww}. This derivation is completely
equivalent to that of Refs.~\cite{CataniHQ,Catani:1993ww}, but it
differs from it because the ladder expansion which underlies the
factorization is treated in a way which is ``dual'' to it in the sense
of Ref.~\cite{duality}: small $x$ logs are factored as part
of standard DGLAP splitting functions, rather than by solving a BFKL~\cite{BFKL}
equation for the gluon Green function. 

The consequence of this is that
the same kinematic as in the proof of collinear
factorization~\cite{CFP} can be used. 
Within this kinematics, the
dependence on transverse and longitudinal momentum components are 
kept separate from each other, and this renders the generalization to
rapidity distributions possible. 
This result is derived in Sect.~\ref{rapidity},
where a simple  formula for the computation of leading log $x$
contributions to rapidity distributions to all orders is arrived at,
in terms of the Fourier-Mellin transform of the partonic rapidity
distribution computed at the lowest nontrivial order with incoming
off-shell gluons.

In the second part of the paper we apply this formalism 
to Higgs production in gluon-gluon fusion. First, in
Sect.~\ref{sec_higgs_res} we summarize the small $x$ behavior of
rapidity distributions in this case, both in
the limit $m_{\rm top}\to\infty$ and for finite top mass, and  
specifically we obtain explicit results for LL$x$ terms
up to NNLO, already known at NLO in the
$m_{\rm top}\to\infty$ limit, but otherwise determined here for the
first time in closed form. 
In
Sect.~\ref{higgsres} we then apply to this case the resummation formula of
Sect.~\ref{rapidity}, by deriving a resummed expression and explicit
coefficients up to NNLO.
 We check that the results up to  $O(\alpha_s^4)$ of the previous
 section are reproduced (which turns out to happen in a rather
 nontrivial way), we  construct an explicit analytic
matched $O(\alpha_s^3)$ 
expression in the finite  $m_{\rm top}$ case,
and compare it to known numerical results~\cite{AnastasiouHiggsNNLO,hpro}. 
We also give an estimate of finite mass corrections at NNLO. Technical
results on $\msb$ factorization and subtraction of collinear
singularities
both at the inclusive
and differential level are collected in two appendices.

\section[High energy factorization  of inclusive cross-sections]{High energy factorization  of inclusive cross - sections}\label{inclusive}

We will rederive high energy
factorization~\cite{CataniHQ,Catani:1993ww} 
in the leading logarithmic approximation (LL$x$)
for inclusive quantities. After some introductory comments, we will
first consider the known case of $n$-gluon emission at the double log 
(i.e. LL$x$-LL$Q^2$) level, and
finally the nontrivial general LL$x$ case. We then recall how the
factorized result can be exploited for resummation, and finally we
compare directly our approach to the original one of
Refs.~\cite{CataniHQ,Catani:1993ww}.
The key idea of our argument is to exploit the factorization of the
partonic cross-section in terms of a hard part and a ladder part, but
then compute the ladder part using standard collinear factorization
and DGLAP evolution, rather than  BFKL evolution, exploiting the fact
that due to perturbative duality~\cite{duality,jaros} they are in fact
equivalent at the leading twist level.

\subsection{Factorized cross-section}

In order to simplify our derivation, we consider at first a  
photo- or lepto-production process
\beq
{\cal V}(n)+g(p)\rightarrow \mathcal{S}+X.
\label{photoprod}
\eeq
characterized by a hard scale $Q^2$. In Eq.~\href{photoprod} $g(p)$
denotes a gluon of momentum $p$,  ${\cal V}(n)$ denotes an
electroweak gauge boson of momentum $n$, not necessarily on-shell:
e.g. for neutral-current Deep-Inelastic Scattering ${\cal V}=\gamma^*$
or ${\cal V}=Z^*$, and $\mathcal{S}$ the desired final state (a quark
for DIS, a dilepton for Drell-Yan and so forth).
Generalization to hadroproduction or
quark-initiated processes is straightforward and will be discussed In
Sect.~\ref{xsres_hadro}.  

\begin{figure}
\begin{center}
\epsfig{file=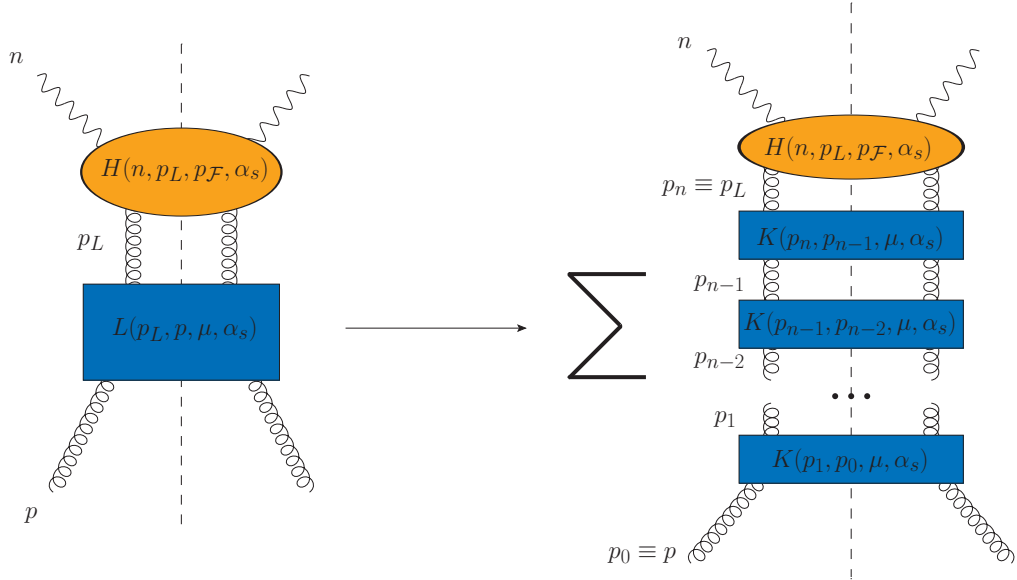, width=0.9\textwidth}
\end{center}
\caption{Left: decomposition of the cut partonic cross-section in terms of
  two-gluon irreducible hard part and a reducible ladder part. Right:
generalized ladder expansion of the ladder part.}
\label{genlad_pic}
\end{figure}

\begin{figure}
\centering
\epsfig{file=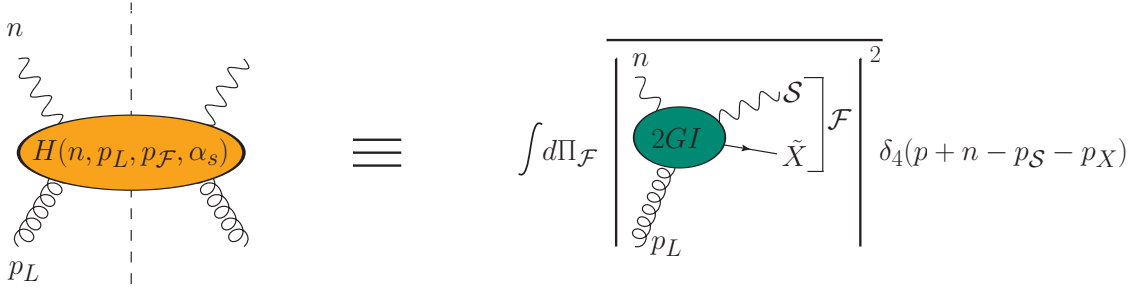, width=1.0\textwidth}
\caption{Graphic representation of the hard part in Eq.~\href{fact2GI}.
Note that the hard part contains the momentum conservation delta function 
as well as the $\mathcal S$ and $\tilde X$ phase space integration, but
it does not contain phase space integration for gluons emitted along the
ladder.}
\label{hard_pic}
\end{figure}

Our starting point is the same of Refs.~\cite{CataniHQ,Catani:1993ww},
namely the observation~\cite{BFKL, CiafaloniCoherence} that in an axial gauge
high energy enhanced terms only come from  cut diagrams which are at least
two-gluon reducible in the $t$-channel (see Fig.~\ref{genlad_pic}). 
The generic dimensionless partonic
cross-section $\sigma$  can then be split in a (process-dependent)
two-gluon irreducible part (``hard'' 
part), and a generally two-gluon reducible part (``ladder'' part):
\be\label{fact2GI}
\sigma = \int \frac{Q^2}{2s}H^{\mu\nu}(n,p_L,p_{\mathcal F},  
\mu_R, \mu_F, \as)\cdot 
L_{\mu\nu}(p_L,p,  \mu_R, \mu_F, \as)  
\left[\d p_L\right],
\ee
where the hard part
$H^{\mu\nu}$ includes the phase space for the observed final state 
$\mathcal S$ and
the momentum conservation delta function and we have also made
explicit the flux factor $1/(2s)$. Here $p_{\mathcal F}$ stands for a set
of momenta parametrizing the final state of the hard part, which we
will henceforth denote as $\mathcal F$, see Fig.~\ref{hard_pic}.
The measure $\left[\d p_L\right]$
is the $p_L$--loop integration measure for the cut diagram 
Fig.~\ref{genlad_pic} and will be specified in the following 
(see Eq.~\href{dpl} below). 

Diagrams like 
Fig.~\ref{genlad_pic} contain in general both ultraviolet and infrared 
divergences, which
must be regulated. This leads to the dependence of the partonic cross-section
on the renormalization $\mu_R$ and factorization $\mu_F$ scales. 
The renormalization scale dependence can be reabsorbed in the running 
coupling $\as(\mu_R)$. Since running coupling effects are NLL$x$, while we
are working at the LL$x$ level, in the following we will omit all the 
$\mu_R$ dependence and consider only the dependence on $\mu_F$, which 
from now on we will call just $\mu$. 

In order to simplify our derivation, we will 
assume further that the hard part is not just 2-gluon irreducible (2GI), but
in fact 2-particle irreducible (2PI). This implies that we can neglect
the $\mu_F=\mu$ dependence in $H^{\mu\nu}$. The generalization 
to the case in which the hard part is 2GI but two-particle
reducible has
been worked out in~\cite{CataniDIS}; the extension of our
formalism to it would be straightforward but we will not discuss it further.
The functions $H^{\mu\nu}$ and $L^{\mu\nu}$ thus depend on the following 
variables:
\bea
&& 
H^{\mu\nu} = H^{\mu\nu} (n,p_L, p_{\mathcal F}, \as) \nonumber \\
&& L^{\mu\nu} = L^{\mu\nu} (p_L, p, \mu, \as) \label{variables_HL}
\eea

Since everything is on-shell and we are working in an axial gauge, we can
decompose both the hard part and the ladder part 
in terms of conserved Lorentz structures time dimensionless scalar
functions:
\bea \label{HL_expansion}
&& H^{\mu\nu}(n,p_L,p_{\mathcal F}, \as)
=\lp-g^{\mu\nu}+\frac{p_L^\mu p_L^\nu}{p_L^2} \rp
H_{\perp}\lp \frac{Q^2}{n\cdot p_L}, \frac{-p_L^2}{Q^2}, 
\Omega_{\mathcal F}, \as \rp + \nonumber \\ 
&&+p_L^2 \lp \frac{p_L^\mu}{p_L^2}-\frac{n^\mu}{n\cdot p_L} \rp
\lp \frac{p_L^\nu}{p_L^2}-\frac{n^\nu}{n\cdot p_L} \rp 
H_{||}\lp \frac{Q^2}{n\cdot p_L}, \frac{-p_L^2}{Q^2}, 
\Omega_{\mathcal F}, \as \rp\label{harddecomp}\\
&& L^{\mu\nu}(p_L, p, \mu, \as)=\frac{1}{p_L^2}
\lp-g^{\mu\nu}+\frac{p_L^\mu p_L^\nu}{p_L^2} \rp
L_{\perp}\lp \frac{-p_L^2}{p\cdot p_L}, \frac{\mu^2}{-p_L^2}, \as\rp \nonumber \\ 
&&+\lp \frac{p_L^\mu}{p_L^2}-\frac{p^\mu}{p\cdot p_L} \rp
\lp \frac{p_L^\nu}{p_L^2}-\frac{p^\nu}{p\cdot p_L} \rp 
L_{||}\lp \frac{-p_L^2}{p\cdot p_L}, \frac{\mu^2}{-p_L^2}, \as\rp
\label{ladderdecomp}
\eea
where $\Omega_{\mathcal F}$ stands for a set of (typically angular, 
see e.g.~\cite{SimoneDY})
 dimensionless variables
which characterize the final state $\mathcal F$. 
Note that we have explicitly extracted 
the propagator factor $1/p_L^2$ from the scalar functions $L_{||,\perp}$.

In the high energy  limit Eq.~(\ref{fact2GI}) simplifies somewhat.
To see this, we work in the center-of-mass frame of the colliding partons
 and introduce the Sudakov parametrization
\be\label{Sudakov}
p_L= z p -k -\frac{k_T^2}{s(1-z)}n=\lp\sqrt{\frac{s}{2}}
  z,-\frac{k_T^2}{\sqrt{2s}(1-z)};-\vec k_T\rp 
\ee
where $k=(0,k_x,k_y,0)$ is a purely transverse four-vector with 
$k^2=-k_T^2<0$ and as usual $s=2 p\cdot n$, and in the last step we
have written the corresponding light-cone components.
 With this parametrization
it is possible to show that the integration measure 
$\left[ dp_L \right]$ for the cut diagram Fig.~\ref{genlad_pic} is
(see e.g.~\cite{CFP})
\be\label{dpl}
\left[ \d p_L \right] = \frac{\d z}{2(1-z)} \d^2 {\bf{k_T}}.
\ee
Since the $z$ integration goes from $x$ to 1, 
we can easily identify the leading small $x$ region with the $z\ll 1$ region. 
As for $k_T^2$, it is typically bounded by the hard scale $Q^2$. Since
in the high energy regime $Q^2\ll s$, we have $k_T^2/s \ll 1$.
Summarizing, at small $x$ we are in the regime
\be\label{sxkin}
z\ll 1; \qquad \frac{k_T^2}{s} \ll 1;
\ee
subleading terms in $z$ and $k_T^2/s$ upon 
integration lead to power-suppressed  $O(x)$ terms. 

This kinematics leads to a simplification in the arguments of the scalar 
functions 
$H_{||, \perp},~L_{||, \perp}$. We have
\bea\label{smallx_arguments}
 H_{||,\perp} 
\lp \frac{Q^2}{n\cdot p_L}, \frac{-p_L^2}{Q^2}, 
\Omega_{\mathcal F}, \as \rp &=& 
H_{||,\perp} 
\lp \frac{Q^2}{z s }, \frac{k_T^2}{Q^2(1-z)}, 
\Omega_{\mathcal F}, \as \rp  \nonumber \\
& =& H_{||,\perp} 
\lp \frac{Q^2}{z s }, \frac{k_T^2}{Q^2}, 
\Omega_{\mathcal F}, \as \rp (1+O(z)) \nonumber \\
L_{||, \perp}\lp \frac{-p_L^2}{p\cdot p_L}, \frac{\mu^2}{-p_L^2}, \as\rp & =&
 L_{||, \perp}\lp -\frac 1 2, \frac{\mu^2}{k_T^2}, \as\rp (1+O(z))
\nonumber \\
&=& L_{||, \perp}\lp \frac{\mu^2}{k_T^2}, \as\rp (1+O(z)).
\eea

With this in mind, and using the fact~\cite{CataniHQ} that the scalar
functions $H_\perp$ and $H_{||}$ have the same small $x$
behavior, and similarly $L_\perp$  and $L_{||}$, 
we can write the differential dimensionless cross-section as
\bea\label{1gluon_fact}
&&\d\sigma = -\frac{Q^2}{2s}\left[
\frac{2}{z^2}
H_{||} \lp \frac{Q^2}{z s},\frac{k_T^2}{Q^2},  \Omega_{\mathcal F}, \as \rp
L_{||}\lp \frac{\mu^2}{k_T^2}, \as \rp\lp 1 + O(z)\rp \right]
\d z \frac{\d^2\bold k_T}{k_T^2} =\nonumber\\
&&=\left[ - \frac{Q^2}{2s z}H_{||} \lp \frac{Q^2}{z s}, \frac{k_T^2}{Q^2}
,\Omega_{\mathcal F}, \as \rp
\lp 1 + O(z) \rp\right]
\left[2\pi L_{||}\lp \frac{\mu^2}{k_T^2},\as\rp\right]
\frac{\d z}{z} \frac{\d k_T^2}{k_T^2}\frac{\d\theta}{2\pi}.
\nonumber\\
\eea
In Ref.~\cite{CataniHQ} the ladder part $L_{\mu\nu}(p_L,p,\mu,\as)$ is
computed at the LL$x$ level in terms of a gluon Green function, which
in turns sums leading logs of $x$ by iterating a BFKL~\cite{BFKL}
kernel.
Here we observe that we may equivalently express $L_{\mu\nu}(p_L,p,\mu,\as)$
in terms of the generalized ladder expansion of Ref.~\cite{CFP}, and
we will thus express it in terms of standard collinear anomalous
dimensions. First, however, we discuss the hard part.

\subsection{Hard part}\label{sec_hard_part}
Let us now concentrate on the process-dependent hard part. We introduce
a hard coefficient function $C$ defined as
\bea\label{coeff}
&& C\lp\frac{Q^2}{zs}, \frac{Q^2}{k_T^2} , \as \rp \equiv
\int_0^{2\pi} \frac{\d \theta}{2\pi} \frac{Q^2}{2 s z} \left[
\mathcal P_{\mu\nu} H^{\mu\nu}(n, p_L, p_{\mathcal F}, \as)\right]
=\nonumber\\
&&
=- 
\int_0^{2\pi}\frac{\d \theta}{2\pi}
\frac{Q^2}{2s z}H_{||} \lp \frac{Q^2}{z s}, \frac{Q^2}{k_T^2}, 
\Omega_{\mathcal F}, \as \rp 
 \lp 1+O(z)\rp,
\eea
where we have defined the projector
\be\label{proj}
\mathcal P^{\mu\nu} \equiv \frac{k^\mu k^\nu}{k_T^2}
\ee
which, up to $O(z)$ terms, selects the
longitudinal part $H_{||}$ of  the full $H^{\mu\nu}$. 
In Eq.~\href{coeff} the explicit dependence of $C$ on $\Omega_{\mathcal F}$ 
is understood.

Equation~\href{coeff} has a simple interpretation: $C$ is the cross-section for the partonic process 
${\cal V}(n) + g^*(q) \rightarrow \mathcal F$ for an off-shell
incoming gluon with momentum 
\be\label{q-off-shell}
q = z p + k; \quad q^2 = -k_T^2.
\ee
In this interpretation, $\mathcal P$ can be thought of as the sum over
the polarizations of the off-shell gluon. Note that 
\be\label{pol_onshell}
\left< \mathcal P^{\mu\nu} \right>_\theta \equiv
\int_0^{2\pi} \frac{d\theta}{2\pi} \mathcal P^{\mu\nu} = \frac{1}{2} 
\lp - g_{\mu\nu} + \frac{q_\mu n_\nu + q_\nu n_\mu}{q\cdot n} \rp
\equiv \frac{d_{\mu\nu}}{2},
\ee
i.e. the azimuthal average of $\mathcal P^{\mu\nu}$ performs the average
over the polarizations of an on-shell gluon with momentum $q$. 
Equations~(\ref{coeff},~\ref{pol_onshell}) then imply
\be\label{onshell}
\lim_{k_T^2 \rightarrow 0} 
\left<C\lp\frac{Q^2}{zs}, \frac{Q^2}{k_T^2},\as \rp\right>_\theta =
\sigma_{\text{on-shell}}\lp {\cal V}(n), g(zp) \rightarrow \mathcal F \rp,
\ee
i.e. in the limit $k_T/Q^2\rightarrow 0$ the 
azimuthal average of the hard coefficient function
$C$ reduces to the on-shell partonic (dimensionless) cross-section.
Note that in high energy kinematics Eq.~(\ref{sxkin}), unlike in
standard collinear kinematics, we cannot assume $k_T^2\ll Q^2$ and thus
the full dependence on $\frac{k_T^2}{Q^2}$ must be retained in the
coefficient function $C$ Eq.~(\ref{coeff}). However, the assumption
that $k_T^2\ll s$ is sufficient~\cite{CataniHQ} 
for the  factorization of the cross-section Eq.~(\ref{fact2GI}) to hold.

For the subsequent discussion, it is important to observe that in the
limit $x\to 0$ the hard part, being 2GI, is finite, i.e. it does not
contain logarithmically enhanced contributions. 
In particular, unless the vertex of the interaction for the process
${\cal V}(n) + g^*(q) \rightarrow \mathcal F$ is pointlike,~\cite{CataniHQ} 
\be\label{limit_resolved}
\lim_{x\to 0} C\lp x, \frac{Q^2}{k_T^2}, \as\rp = 0.
\ee
This is for instance the case for Deep-Inelastic Scattering and for
 heavy quark, Drell-Yan or Higgs
production with finite $m_{\rm top}$. For Higgs production in the
$m_{\rm top}\to\infty$ limit the effective gluon-gluon-Higgs
interaction is pointlike, and in such a case~\cite{HautmannHiggs} 
\be\label{limit_renormalizable}
\lim_{x\to 0} C\lp x, \frac{Q^2}{k_T^2}, \as\rp = C_0
\ee
with $C_0$ some constant. As it is well  known, and as we will see
explicitly in Sect.~\ref{sec_higgs_res}
the 
behavior Eq.~(\ref{limit_renormalizable}) leads to double logarithmic
small~$x$ singularities. 

\subsection{Ladder part: generalities and double log approximation}
\label{genladdl}

We now turn to the computation of the ladder part. Unlike the hard
part, which is 2PI, the ladder part may contain collinear
singularities. In general, the collinear singularities of the partonic 
cross-section $\sigma$, Eq.~(\ref{fact2GI}), originate from those 
of the ladder part, and from the
collinear region of the integration of the momentum $p_L$ which
connects the hard and ladder part. Collinear singularities can be
subtracted from the partonic cross-section and factored in the parton
distributions, to which the hard cross-section is connected by the
lower line with momentum $p$, using the iterative procedure of
Ref.~\cite{CFP}, based on a generalized ladder expansion. 
As already mentioned, after
regularization and factorization of the collinear singularities the
ladder part is a function of $\frac{k_T^2}{\mu^2}$, where $\mu$ is the
(factorization) scale introduced in the course of regularization.

In the generalized ladder expansion,  $L_{||}$
is  written as a sum of terms  each of which, after
regularization and factorization, leads to a contribution proportional to
$\ln^{n}\frac{k_T^2}{\mu^2}$ to the cross-section, with $n$ ranging
from one to infinity. 
The $n$-th order contribution comes from a ladder
diagram obtained iterating $n$ times a kernel $K(p_i,p_{i-1},\mu, \as)$ with
$i=1,2,\dots,n$. To simplify our notation, from now on 
we will make the identifications $p_n\equiv p_L$, $p_0\equiv p$, see 
Fig.~\ref{genlad_pic}. 
Subsequent
iterations of the kernel are connected by a pair of $t$-channel lines
with momenta $p_n$ (see Fig.~\ref{genlad_pic}) and ordered transverse
momenta
$k^2_{T,\,1}\ll k^2_{T,\,2}\ll \dots\ll k^2_{T,\,n} = k^2_{T}$. Because we
are interested in the high energy limit, reducible lines 
are all gluon lines: quark contributions only appear at NLL$x$ and
subsequent orders~\cite{CataniHQ}.
While we refer to Ref.~\cite{CFP} for a detailed discussion, the way
the factorization of collinear singularities is performed in the
generalized ladder expansion is summarized in Appendix~\ref{app_CFP}.

We will
now show how to compute the ladder part using the generalized ladder
expansion in the double logarithmic approximation, i.e. in the very
simple case in
which the result is determined simultaneously at the LL$x$ and
LL$Q^2$, and then turn in
Sect.~\ref{llxladder} to the more general LL$x$ 
case which we are interested in. At LL$x$-LL$Q^2$, 
the kernel $K(p_i,p_{i-1}, \mu, \as)$ 
must be
linear in $\alpha_s$ because the highest power of 
$\ln^{n}\frac{k_T^2}{\mu^2}$  in the cross-section
must coincide with the order in $\alpha_s$. We define thus 
\beq
K^1 \lp \frac{\mu^2}{k_T^2},\as\rp \equiv 
K^{{\rm LL}Q^2}_{||}\lp \frac{\mu^2}{k_T^2},\as\rp 
\label{llqk}
\eeq
with $K^1$ linear in $\alpha_s$.

Consider first the case in which the kernel $K^1 $ is only inserted
once. The corresponding (integrated) contribution to the
cross-section Eq.~(\ref{1gluon_fact}) is given by
\be\label{1gluon_xspace}
\bar\sigma^1\lp \frac{Q^2}{s},\frac{\mu^2}{Q^2}, \as\rp 
= \int_x^1 \frac{\d z}{z} \int \frac{\d k_T^2}{k_T^2}
C\lp \frac{Q^2}{ z s}, \frac{Q^2}{k_T^2},\as\rp
\left[2\pi K^1\lp \frac{\mu^2}{k_T^2}, \as \rp\right],
\ee
where the bar over $\sigma$ indicates 
that collinear singularities have still to be
subtracted from the cross-section, so this expression only makes sense at a regularized level.
It is convenient to introduce the dimensionless variables 
\be\label{x_xi_defs}
x\equiv \frac{Q^2}{s};\qquad    \xi \equiv \frac{k_T^2}{Q^2},
\ee
in terms of which Eq.~\href{1gluon_xspace} becomes
\be\label{1gluon_x_xi}
\bar\sigma^1\lp x,\frac{\mu^2}{Q^2}, \as\rp 
= \int_x^1 \frac{\d z}{z} \int \frac{\d \xi}{\xi}
C\lp \frac{x}{ z }, \xi,\as\rp
\left[2\pi K^1\lp \frac{\mu^2}{Q^2 \xi}, \as \rp\right].
\ee
Note that in terms of these variables
\be\label{mu2/(q2 xi)}
\frac{\mu^2}{Q^2 \xi} = \frac{\mu^2}{k_T^2},
\ee
i.e. the combination $Q^2\xi/\mu^2$
is $Q^2$ independent, and  a 
$\ln^n \frac{k_T^2}{\mu^2}$ term shows up as 
$\ln^n \frac{Q^2\xi}{\mu^2}$. 

The $z$ convolution in Eq.~(\ref{1gluon_x_xi}) is turned into an
ordinary product by Mellin transformation:
\be\label{nmellin}
f(N)\equiv \int_0^1 \d x x^{N-1} f(x);\qquad 
f(x)=\int_{c-i\infty}^{c+i\infty} \frac{\d N}{2\pi i} x^{-N}f(N)
\ee
(note that by slight abuse of notation we denote with the same letter
both the function and its transform).
We get
\be\label{1gluon_4d}
\bar\sigma^1\lp N,\frac{\mu^2}{Q^2},\as\rp
= \int\frac{\d \xi}{\xi}
C\lp N, \xi,\as\rp
\left[\frac{2\pi }{N}K^1\lp \frac{\mu^2}{Q^2\xi},\as\rp
\right].
\ee
Since in the limit $\xi\to 0$ (i.e. $k_T^2\rightarrow 0$ )
the coefficient function $C$ does
not vanish (see Eq.~\href{onshell}), Eq.~\href{1gluon_4d} is collinear
divergent. We  use dimensional regularization and let
$d=4-2\epsilon$. 
It can be shown~\cite{CataniDIS} that in $d$ dimensions
$\mathcal P^{\mu\nu}$ is still given by Eq.~\href{proj}, hence 
Eq.~\href{onshell} remains true also in $d=4-2\epsilon$ dimensions. 

The regularized expression is explicitly given 
by\begin{footnote}{Henceforth, to simplify notation we omit
terms like 
$(4\pi)^\epsilon/\Gamma(1-\epsilon)$, which  in the $\msb$ scheme are
  subtracted anyway. Full details on the calculation 
are given in  Appendix~\ref{app_CFP}.}\end{footnote}:
\be\label{1g_d}
\bar\sigma^1\lp N, \frac{\mu^2}{Q^2}, \as; \epsilon\rp
\int_0^\infty \frac{d\xi}{\xi^{1+\epsilon}}
C\lp N, \xi, \as; \epsilon \rp
\left[
\frac{2\pi }{N}
K^1\lp  \lp\frac{\mu^2}{Q^2}\rp^\epsilon\as;N, \epsilon\rp\right],
\ee
where the dependence of $K^1$ on $N$ is $O(\epsilon)$.
In Eq.~\href{1g_d} the limit $\xi\rightarrow \infty$ is never
reached, because $k_T^2$ is bounded by the kinematics 
(typically $k_T^2<Q^2$). Note that we have
used the assumption that the hard part
is 2PI, which ensures that $C$ is free of collinear 
singularities~\cite{CFP, EllisGeorgi}. The generalization to processes
for which $H$ is 2GI but not 2PI (such as  Deep-Inelastic
Scattering and Drell-Yan) requires also treating the collinear
singularities in $H$~\cite{CataniDIS, SimoneDY}. Note that the
dependence on $\mu^2$ enters only
through the combination
 $ (\mu^2)^\epsilon\as$ and it is thus fixed by dimensional analysis.
 
Using the expansion
\be\label{epsilon_expansion}
\frac{1}{\xi^{1+\epsilon}} = -\frac{\delta(\xi)}\epsilon 
+ \sum_{i=0}^{\infty} \left[\frac{\ln^i \xi}{\xi} \right]_+
\frac{(-\epsilon)^i}{i!}
\ee
in  Eq.~\href{1g_d} we get
\be\label{coll_limit}
\bar\sigma^1\lp N, \frac{\mu^2}{Q^2}, \as; \epsilon\rp=
-\frac{1}{\epsilon}
\sigma_{\text{on-shell}} \times \left[
\frac{2\pi}{N}K^1\lp \as \rp\right] + \text{finite},
\ee
where $\sigma_{\text{on-shell}}$ is the on-shell cross-section for
the process ${\cal V}(n)+g(zp) \rightarrow \mathcal F$, see 
Eq.~\href{onshell}.
Equation~\href{coll_limit} shows that $2\pi  K^1(\as)/N$ 
is the residue of the $O(\alpha_s)$
collinear pole, i.e. the leading--order anomalous dimension. Because
all the computation has been performed at the LL$x$ level, only the
leading small $x$ contribution to the anomalous dimension is significant,
so
\bea\label{K1_g0}
&&\frac{2\pi}{N}K^1(\as) = 
\asb \gamma_0(N)\nonumber;\qquad  \asb\equiv\frac{\as N_c}{\pi}\\
&&\qquad \gamma_0(N)=1/N.
\eea

Having determined $K^1$, we can  
substitute  it in Eq.~\href{1g_d}:
\be\label{1g_final}
\bar\sigma^1 \lp N, \frac{\mu^2}{Q^2}, \as; \epsilon\rp = 
\int_0^{\infty} \frac{d\xi}{\xi^{1+\epsilon}}
C\lp N, \xi, \as; 
\epsilon\rp
\asb\lp \frac{\mu^2}{Q^2}\rp^\epsilon \gamma_0(N),
\ee
where we used the fact that the LO gluon-gluon anomalous dimension is
$\epsilon$ independent. 
Equation~\href{1g_final} is the  expression for
the regularized (collinear singular) 
cross-section ${\cal V}(n)+g(p)\rightarrow \mathcal F + g$ 
in the small $x$ limit.

We are now ready to compute the ladder expansion of $L$ at LL$x$, LL$Q^2$.
The regularized contribution to the cross-section when the kernel is
iterated $n$ times is
\bea\label{ngluon_kt}
&& \bar\sigma^n\lp N, \frac{\mu^2}{Q^2}, \as; \epsilon\rp = 
\left[\asb\lp\frac{\mu^2}{Q^2}\rp^\epsilon \gamma_0(N)\right]
\int_0^\infty \frac{d \xi_n}{\xi_n^{1+\epsilon}}
C\lp N, \xi_n, \as; \epsilon\rp
\times \nonumber\\
&& \times
\int_0^{\xi_n} \left[\asb\lp\frac{\mu^2}{Q^2}\rp^\epsilon \gamma_0(N)\right]
\frac{d\xi_{n-1}}{\xi_{n-1}^{1+\epsilon}}\times...
\times\int_0^{\xi_2}\left[\asb\lp\frac{\mu^2}{Q^2}\rp^\epsilon \gamma_0(N)\right]
\frac{d\xi_{1}}{\xi_{1}^{1+\epsilon}}.
\eea
Had we used a collinear 
cutoff $\mu_F$ instead of dimensional regularization,
the ordered integrations
in Eq.~\href{ngluon_kt} would give rise to the expected $\ln^{n} Q^2/\mu_F^2$
term. In dimensional regularization,  Eq.~\href{ngluon_kt} contains 
collinear $\epsilon$ poles. Their $\msb$
factorization using the generalized
ladder expansion of Ref.~\cite{CFP} is performed by
requiring Eq.~\href{ngluon_kt} to be
finite after each $\xi_i$ integration. This is done subtracting iteratively
in such a way that Eq.~\href{ngluon_kt} is finite also at the 
integrand level; details are given in
Appendix~\ref{app_CFP}. This then leaves a single $n$-th order
$\epsilon$ pole in the cross-section, which can be subtracted using
standard $\msb$.

The result after iterative subtraction of the first $n-1$
singularities is
\bea\label{ngluon_msb}
&&\sigma^n\lp N, \frac{\mu^2}{Q^2}, \as; \epsilon\rp =
\left[\asb\lp\frac{\mu^2}{Q^2}\rp^\epsilon \gamma_0(N)\right]
\times \nonumber\\
&&\quad\times
\int_0^\infty \frac{d \xi_n}{\xi_n^{1+\epsilon}} 
C\lp N, \xi_n, \as; \epsilon\rp
\frac{1}{(n-1)!}\frac{1}{\epsilon^{n-1}}
\left[\asb \gamma_0(N)
\lp1-\lp \frac{\mu^2}{Q^2 \xi_n} \rp^\epsilon\rp
\right]^{n-1},
\eea
which contains a single  $1/\epsilon^n$ pole, as it ought to 
in the $\msb$ scheme.

The full LL$x$-LL$Q^2$ result is found by adding up all contributions
Eq.~(\ref{ngluon_msb}). It turns out that (as well known from
Ref.~\cite{CataniHQ} for the general LL$x$ case) the sum is actually
finite, i.e. the exponentiation of the leading collinear singularities
to all orders leads effectively to a cross-section evaluated
with off-shell gluons,  which is thus finite.
Indeed, we get
\be\label{kt_res}
\sigma = \sum_{n=1}^\infty \sigma^n = 
\left[\asb \lp \frac{\mu^2}{Q^2}\rp^\epsilon\gamma_0(N)\right]
\int_0^\infty \frac{d\xi}{\xi^{1+\epsilon}}
C\lp N, \xi, \as; \epsilon\rp
e^{
\asb \gamma_0(N)
\frac{1}{\epsilon} 
\left[ 1-\lp\frac{\mu^2}{Q^2\xi}\rp^\epsilon\right]},
\ee
which is finite in the $\epsilon\rightarrow 0$ limit. Indeed
\be\label{deq4_limit}
\lim_{\epsilon\rightarrow 0} \frac{1}{\xi^{1+\epsilon}}
\exp\left[
\asb  \gamma_0(N)
\frac{1}{\epsilon} 
\lp 1-\lp\frac{\mu^2}{Q^2\xi}\rp^\epsilon\rp
\right] = \xi^{\asb \gamma_0(N)-1}
\exp\left[
\as \gamma_0(N) \ln \frac{Q^2}{\mu^2}
\right],
\ee
and since $\asb \gamma_0(N) > 0$ also the last collinear divergence in 
Eq.~\href{kt_res} is regulated.

We can then safely go back to four dimensions. For completeness we 
reintroduce the argument of the running coupling $\as$, although 
running coupling is a NLL$x$ effect. We obtain
\bea\label{g0_iter_mu}
&&\sigma\lp N, \frac{\mu^2}{Q^2}, \as(\mu^2)\rp = 
\asb(\mu^2) \gamma_0(N) \times \nonumber\\
&& \quad \times
\int_0^\infty d\xi \xi^{\asb(\mu^2) \gamma_0(N) -1}
e^{\asb(\mu^2) \gamma_0(N) \ln\frac{Q^2}{\mu^2}}
C\lp N, \xi, \as(\mu^2)\rp,
\eea
or, using renormalization group invariance to set 
$\mu^2=Q^2$, thereby absorbing the explicit $\mu^2/Q^2$ dependence
into the 
running coupling $\as(Q^2)$,
\be\label{g0_iter}
\sigma\lp N, \as(Q^2)\rp = 
\asb(Q^2) \gamma_0(N)
\int_0^\infty d\xi \xi^{\asb(Q^2) \gamma_0(N) -1}
C\lp N, \xi, \as(Q^2)\rp,
\ee
where we have made explicit the fact that the partonic cross-section no longer
depends on $\mu^2/Q^2$. 
At each fixed order in $\alpha_s$
 Eq.~\href{g0_iter} reproduces the small $N$ limit of the coefficient function, 
as can be checked by integrating by parts and expanding in powers of $\as$.

We see that there are 
two sources for the $N$--dependence of the cross-section $\sigma$: the
$\asb \gamma_0(N)$ terms and the explicit $N$--dependence of $C$. 
Let us first assume that $C(N,\xi,\as)$ is regular in $N=0$ (i.e. 
$C(x,\xi,\as)\to 0$ for $x\to 0$, see the discussion at the end 
of Sec.~\ref{sec_hard_part}):
\be\label{good_N_behavior}
C(N, \xi, \as) = C(0,\xi,\as) + N C'(0,\xi,\as) + ...
\ee
In this case Eq.~\href{g0_iter} can be written as (see Eq.~\href{K1_g0})
\bea\label{g0_iter_nospurious}
\sigma(N,\as) &=& \frac{\asb}{N}\int \d \xi \xi^{\frac{\asb}{N}-1}
C(0,\xi,\as) (1 + O(N))\nonumber\\
&=&\as^k \sum_{i=0}^{\infty} c_i^{\text{LL}x-\text{LL}Q^2} 
\lp\frac{\asb}{N}\rp^i (1+O(N)),
\eea
where the power $k$ is process-dependent and is encoded in $C$ (e.g. 
$k=1$ for Drell-Yan and DIS, $k=2$ for Higgs). 
The expansion in $\as/N$ leads to the desired $\as \ln \frac 1 x$ 
expansion in 
$x$ space. From Eq.~\href{g0_iter_nospurious} it is then clear that only
the $N$--independent term $C(0,\xi,\as)$ is relevant for a LL$x$ 
calculation. 
Things are different when $C(x,\xi,\as)\to C_0\ne 0$ for $x\to 0$. 
This is for instance the case for the Higgs coefficient function in the 
infinite $m_{\rm top}$ approximation~\cite{HautmannHiggs}. In this
case $C(N,\xi,\as)$ is not regular in $N=0$, and the full $N$--dependence 
must be kept. The singular behavior of the coefficient function in
$N=0$ leads to double log behavior in $x$ space, as we shall see explicitly in 
Sec.~\ref{higgsres}.

Eq.~\href{g0_iter} reproduces the LL$x$ result of Ref.~\cite{CataniHQ} in
the LL$Q^2$ case:  the LL$x$ coefficient function is found
evaluating the partonic cross-section with incoming off-shell gluon,
performing a Mellin transform with respect to the gluon virtuality,
and then letting $M=\gamma$, where $M$ is the Mellin variable which is
conjugate to the gluon virtuality, and $\gamma$ is the relevant
anomalous dimension, which in the LL$Q^2$ case 
coincides with the small $x$ limit of the leading
order anomalous dimension $\gamma_0$.

\subsection{Ladder part: LL$x$ case}
\label{llxladder}

The argument presented in Sect.~\ref{llxladder} actually also works in the
general LL$x$ case, at the price of some technical complication.
In this case the kernel $K$ contains contributions to all orders in
the strong coupling $\alpha_s$. The way the ladder expansion is
constructed when the kernel $K$ is defined at a generic order in
$\alpha_s$ was explained in Ref.~\cite{CFP}, and it is that which is
used to compute anomalous dimensions at any perturbative order. In a
nutshell, the ladder diagram of Fig.~\ref{genlad_pic} is still
constructed with strongly ordered transverse momenta in such a way
that all large logs of $Q^2$ arise due to integration over the
momenta $p_i$. Unordered contributions, as well as finite parts due to
iterations of lower--order kernel after subtraction of collinear poles,
are included as higher order contributions to the kernel $K$, which
is now no longer 2PI, even though it is still free of collinear
singularities. 
The total cross-section is still the sum of
contributions with $n$ iterations of the kernel, with $n$ ranging from
one to infinity.

With a generic kernel,  the steps leading to Eq.~\href{1g_d}
remain unchanged, and lead to
\bea\label{ng_d}
&&\bar\sigma^1\lp N, \frac{\mu^2}{Q^2}, \as; \epsilon\rp 
=\nonumber\\
&=&\int_0^\infty \frac{d\xi}{\xi^{1+\epsilon}}
C\lp N, \xi, \as; \epsilon\rp
K\lp N, \lp\frac{\mu^2}{Q^2\xi}\rp^\epsilon, \as; \epsilon\rp
\eea
Because $K$ is free of collinear singularities, the analogue
of Eq.~\href{coll_limit} also still holds:
\beq\label{coll_limit_LLx}
\bar\sigma^1\lp N, \frac{\mu^2}{Q^2}, \as; \epsilon\rp 
=
-\frac{1}{\epsilon} 
\sigma_{\text{on-shell}}
K\lp N,1,\as\rp + \text{finite},
\eeq
which again implies that $K$ is just the standard 
anomalous dimension:
\be\label{genandim}
K\lp N, \lp\frac{\mu^2}{Q^2\xi} \rp^\epsilon, \as; \epsilon\rp =
\gamma\lp N, \lp\frac{\mu^2}{Q^2\xi} \rp^\epsilon,\as ; \epsilon\rp. 
\ee
At the LL$x$ level at which we are working $\gamma$ is the ($d$-dimensional)
LL$x$  anomalous dimension.

However, unlike in the LL$Q^2$ case, the anomalous
dimension $\gamma$ now explicitly depends on $\epsilon$; this
affects the way the  $\msb$ subtraction is performed.
 Indeed Eq.~\href{ngluon_kt} now becomes
\bea\label{nkernel_kt}
\bar\sigma^n\lp N, \frac{\mu^2}{Q^2}, \as;  \epsilon\rp = 
\int_0^\infty
\left[\gamma\lp N, \lp\frac{\mu^2}{Q^2\xi_n} \rp^\epsilon, \as; \epsilon\rp\right]
 \frac{d \xi_n}{\xi_n^{1+\epsilon}} 
C\lp N, \xi_n, \as; \epsilon\rp
\times \nonumber\\
\times
\int_0^{\xi_n} 
\left[\gamma\lp N, \lp\frac{\mu^2}{Q^2\xi_{n-1}} \rp^\epsilon, \as; \epsilon\rp\right]
\frac{d\xi_{n-1}}{\xi_{n-1}^{1+\epsilon}}\times...
\times\int_0^{\xi_2}
\left[\gamma\lp N, \lp\frac{\mu^2}{Q^2\xi_1} \rp^\epsilon, \as; \epsilon\rp\right]
\frac{d\xi_{1}}{\xi_{1}^{1+\epsilon}},\nonumber\\
\eea
which, after iterative subtraction of the first $n-1$
singularities becomes
\bea\label{nkernel_msb}
\sigma^n\lp N, \frac{\mu^2}{Q^2}, \as;  \epsilon\rp =
\gamma\lp N, \lp\frac{\mu^2}{Q^2} \rp^\epsilon, \as; \epsilon\rp
\int_0^\infty \frac{d \xi_n}{\xi_n^{1+\epsilon}} 
C\lp N, \xi_n, \as; \epsilon\rp
\times \nonumber\\
\times
\frac{1}{(n-1)!}\frac{1}{\epsilon^{n-1}}
\left[
\sum_i \frac{\tilde \gamma_i(N,\as;0)}{i}
\lp 1 - \lp \frac{\mu^2}{Q^2 \xi_n} \rp^{i\epsilon}
\frac{\tilde \gamma_i(N,\as; \epsilon)}{\tilde \gamma_i(N,\as;0)}\rp
\right]^{n-1},
\eea
where we have introduced the expansion
\be\label{andim_mu2exp}
\gamma\lp N, \lp\frac{\mu^2}{Q^2\xi}\rp^\epsilon, \as; \epsilon\rp 
= \sum_{i=0}^\infty \tilde\gamma_i\lp N, \as; \epsilon\rp 
\lp\frac{\mu^2}{Q^2\xi}\rp^{i \epsilon}.
\ee
If $\gamma$ is given at the $k$-th--perturbative order
the sum goes from 0 to $k$. 

Again, the total cross-section is rendered finite by exponentiation of
the collinear poles:
\bea\label{kt_resfull}
&&\sigma = \sum_{n=1}^\infty \sigma^n = 
\gamma\lp N, \lp\frac{\mu^2}{Q^2} \rp^\epsilon, \as; \epsilon\rp
\int_0^\infty \frac{d\xi}{\xi^{1+\epsilon}}
C\lp N, \xi, \as; \epsilon\rp \times \nonumber\\
&&\quad
\times
\exp\left[
\frac{1}{\epsilon}\sum_i \frac{\tilde \gamma_i(N,\as; 0)}{i}
\lp 1 - \lp \frac{\mu^2}{Q^2 \xi} \rp^{i\epsilon}
\frac{\tilde \gamma_i(N,\as; \epsilon)}{\tilde \gamma_i(N,\as; 0)}\rp
\right].
\eea
However, we must keep track of the $\epsilon$ dependence in
$\gamma_i$. Expanding
\be
\tilde \gamma_i(N,\as;\epsilon) \equiv \tilde \gamma_i(N,\as) 
+ \epsilon \dot {\tilde \gamma}_i(N,\as) + 
\epsilon^2 \ddot {\tilde \gamma}_i(N,\as) + ...
\label{andimexp}
\ee
we get
\bea
&&\lim_{\epsilon\rightarrow 0} \frac{1}{\xi^{1+\epsilon}}
\exp\left[
\frac{1}{\epsilon}\sum_i \frac{\tilde \gamma_i(N,\as;0)}{i}
\lp 1 - \lp \frac{\mu^2}{Q^2 \xi} \rp^{i\epsilon}
\frac{\tilde \gamma_i(N,\as;\epsilon)}{\tilde \gamma_i(N,\as;0)}\rp
\right]=\nonumber\\
&&\quad =
\xi^{\gamma(N,\as)-1}\exp\left[\gamma(N,\as)\ln\frac{Q^2}{\mu^2}\right]
\mathcal R(N,\as),\nonumber\\
 &&\quad \mathcal R(N,\as)\equiv
\exp\left[ -\sum_i \frac{\dot {\tilde\gamma}_i(N,\as)}{i}\right].
\label{finresgen}
\eea

In comparison to the LL$Q^2$ result Eq.~(\ref{kt_res}) 
the process independent but
scheme dependent 
$\mathcal R$ factor has now appeared. Note that,
 because we are working
in the LL$x$ approximation,  $\as$ does not run. At NLL$x$ and beyond
the running of the coupling generates further contributions to
$\mathcal R$, that depend on higher order terms in the expansion~(\ref{andimexp}) in powers of $\epsilon$~\cite{Ciafaloni:2005cg, SimoneBFKL}.
The LL$x$ cross-section is finally
\be\label{l_iter0}
\sigma\lp N, \as(Q^2)\rp = 
\gamma\lp \frac{\as(Q^2)}{N} \rp
\int_0^\infty d\xi \xi^{\gamma\lp \frac{\as(Q^2)}{N} \rp -1}
C\lp N, \xi, \as(Q^2)\rp \mathcal R\lp\frac{\as(Q^2)}{N}\rp,
\ee
where without loss of generality we have set $\mu^2=Q^2$ (thus omitting
the explicit dependence on $\mu^2/Q^2$) and we have
noticed that the LL$x$ anomalous dimension which enters
Eq.~(\ref{l_iter0}) only depends on $\alpha_s$ and $N$ through the
combination $\frac{\alpha_s}{N}$. As in the LL$Q^2$ case, if 
$C\lp N,\xi,\as(Q^2)\rp$ is regular in $N=0$, then
 only $C\lp 0, \xi, \as(Q^2)\rp$ is relevant at LL$x$ (see discussion
around Eq.~\href{good_N_behavior}).

However,  there is a further source of scheme dependence of our result.
Indeed, we have seen that after exponentiation the last collinear
singularity disappears; it is however present if the result is
expanded out and determined order by order in perturbation theory. It
is thus not obvious that the $\msb$ subtraction commutes with 
exponentiation, and therefore the result  Eq.~\href{finresgen} might
be defined in  a scheme which differs from $\msb$ by a further LL$x$
scheme change, which amounts to a redefinition of parton distributions
by a factor $\mathcal
N(\as/N)$~\cite{Ball:1995tn,Ciafaloni:1995bn,Altarelli:2000mh}.

Hence, our result has the general form
\be\label{l_iter}
\sigma\lp N, \as(Q^2)\rp = 
\gamma\lp \frac{\as(Q^2)}{N} \rp
\int_0^\infty d\xi \xi^{\gamma\lp \frac{\as(Q^2)}{N} \rp -1}
C\lp N, \xi, \as(Q^2)\rp R\lp\frac{\as(Q^2)}{N}\rp,
\ee 
with
\be\label{r_def}
R\lp \frac{\as}{N}\rp \equiv \mathcal N\lp \frac{\as}{N}\rp 
\mathcal R\lp \frac{\as}{N}\rp.
\ee
The factor $\mathcal N$ (and thus $R$) which fixes the factorization
scheme  has been determined
for $\msb$ through a computation of the LL$x$
gluon Green function~\cite{Catani:1993ww}.

\subsection{Cross-section resummation}
\label{xsres}

The LL$x$ cross-section Eq.~\href{l_iter} can be written 
in terms of the
$M$--Mellin transform of 
the coefficient function $C$
\be
h(N,M,\as) \equiv
M 
\int_0^\infty d\xi \xi^{M-1}
C\lp N, \xi, \as \rp R_{\msb}\lp M \rp,
\label{impfac}
\ee
supplemented by the condition
\be
M = \gamma\lp \frac \as N \rp,
\label{mfixing}
\ee
namely
\be
\sigma\lp N, \as\rp
= h\lp N,\gamma\lp \frac \as N \rp,\as\rp. 
\label{cataniresum}
\ee
Explicit resummed expressions can be obtained from
Eq.~(\ref{cataniresum}) 
by substituting
in it the well-known~\cite{BFKL,jaros} expression for the LL$x$
anomalous dimension 
\be\label{gammas}
\gamma\lp \frac \as N \rp =  
  \frac \asb N +2 \zeta (3) \lp  \frac \asb N \rp^4+2 \zeta (5) 
\lp \frac \asb N \rp^6+12 \zeta (3)^2 \lp \frac \asb N \rp^7+...
\ee
and for $R$ the appropriate to the factorization scheme.

The result~(\ref{impfac}-\ref{mfixing}) coincides with that
of Ref.~\cite{CataniDIS}, 
where it was derived by solving the small $x$ evolution equation for 
the gluon Green function. Indeed, our derivation parallels that of
Ref.~\cite{CataniDIS}, the difference being that we have determined
the ladder part using the generalized ladder expansion of
Ref.~\cite{CFP}, while in Ref.~\cite{CataniDIS} it was computed as the
gluon Green function. These coincide because of DGLAP-BFKL
duality~\cite{duality}. 

Explicitly, our Eq.~\href{1gluon_fact} in Ref.~\cite{CataniDIS} is written
in the form
\be\label{CataniFact}
\sigma = \int \frac{dz}{z}\frac{d  k_T^2}{ k_T^2}
C\lp \frac{x}{z}, \frac{k_T^2}{Q^2}, \as\rp 
\mathcal G  \lp z, k_T^2 \rp,
\ee
where  $\mathcal G$ is a gluon Green function, which satisfies a LL$x$
(BFKL) evolution equation and $C$ is defined as in Eq.~\href{coeff}.
Taking an $M$-Mellin transform
Eq.~(\ref{impfac}) both of the coefficient function $C$ and the Green
function $\mathcal G$, and noting that in $M$ space the Green function
has the form
\be
\mathcal G(N,M) =\frac{r(M)}{M-\gamma(\as/N)}\left[1 + O(M-\gamma(\as/N))\right],
\label{greenfunc}
\ee
one ends up with the expression
\beq
\sigma (N,\as)= \int_{c-i\infty}^{c+i\infty}
\frac{dM}{2\pi i} \lp \frac{\mu^2}{Q^2} \rp^{-M} h(N,M) \mathcal G(N,M) 
\label{impfacconv}
\eeq
of the resummed cross-section, which is immediately seen to coincide
with Eq.~\href{l_iter} (with $\mu^2=Q^2$), 
by performing the $M$ integral and
identifying the $R$ function with 
\be
R(\as/N)=r(\gamma(\as/N)).
\label{rfacm}
\ee

The form~(\ref{CataniFact}) of the resummed result has the
structure of a high energy factorization theorem. It is important
however to note that $\sigma$ on the left-hand side of
Eq.~(\ref{CataniFact}) is a partonic cross-section. Namely, the basic
insight of Ref.~\cite{CataniHQ} is that, after factoring the hadronic
cross-section into  parton distributions convoluted with a partonic
cross-section, at the LL$x$ level the partonic cross-section can in
turn be itself factorized into a hard part and a ladder part. The
argument presented in this section has pushed one step further 
the analogy between the
factorization of the hadronic process and the subsequent factorization
of the partonic cross-section, by using the standard collinear
factorization technique of Ref.~\cite{CFP}, albeit at the LL$x$ level,
to compute the ladder part of this second factorization.

\subsection{Hadroproduction and quark-initiated processes}
\label{xsres_hadro}

We now discuss briefly how our results can be generalized to the case
of hadroproduction
processes. This differs from the case discussed so far because there
are now two incoming partons. Consider the process
$g(n)+g(p)\rightarrow \mathcal{S}+X.$
 We have now two different transverse momenta $k_T$, one for each
 leg. 
In principle, the interference between the two legs could spoil
the factorized result Eq.~\href{fact2GI}; however, these contributions do
not appear at LL$x$~\cite{CataniHQ}, so the result in this case is
simply found by applying the generalized ladder expansion to both
legs. In particular, Eq.~\href{fact2GI} generalizes to 
\be\label{fact2GI_hadro}
\sigma \equiv \int \frac{Q^2}{2s}H^{\mu\nu\bar\mu\bar\nu}(n_L,p_L,p_{\mathcal F},  
\as)\cdot 
L_{\mu\nu}(p_L,p,\mu  , \as)  
L_{\bar\mu\bar\nu}(n_L,n,  \mu, \as)  
\left[\d p_L\right] \left[\d n_L\right],
\ee
where again we have assumed that $H$ is 2PI, $\mu$ is the factorization 
scale and we have omitted any dependence on the renormalization scale. 
We have assigned bar index to contributions from the upper leg. 
In Eq.~\href{fact2GI_hadro} $n_L$ is defined through the Sudakov 
parametrization (see Eq.~\href{Sudakov} for comparison)
\be\label{Sudakov_n}
n_L=\bar z n - \bar k - \frac{\bar k_T^2}{s(1-\zbar)} p,
\ee
with $\bar k = (0, \bar k_x, \bar k_y)$ and $\bar k ^2 = -\bar k_T^2 <0$. In 
this parametrization we have (see Eq.~\href{dpl})
\be\label{dnl}
\left[d n_L\right] = \frac{d\zbar}{2(1-\zbar)}d^2 {\bf{\bar k_T}}^2=
\frac{d\zbar}{4(1-\zbar)}d \bar k_T^2 d \bar \theta
\ee

In the high energy regime we have 
$\zbar \ll 1$, $\frac {\bar k_T^2}{s} \ll 1$ (see~\href{sxkin}). Given the factorized form
Eq.~\href{fact2GI_hadro}, we can define a hadroproduction coefficient
function $C$:
\be\label{coeff_hadro}
C\lp \frac{x}{z\zbar}, \xi, \bar \xi, \as\rp \equiv
\int_0^{2\pi} \frac{d\theta}{2\pi}\frac{d\bar\theta}{2\pi}
\frac{Q^2}{2 s z \zbar} \mathcal P_{\mu\nu} \bar{\mathcal P}_{\bar\mu\bar\nu}
H^{\mu\nu\bar\mu\bar\nu}(n_L,p_L,p_{\mathcal F},\as),
\ee
where again we have omitted the $\Omega_{\mathcal F}$ dependence and
\be\label{xi_proj_hadro}
\bar \xi \equiv \frac{\bar k_T^2}{Q^2}; \qquad 
\bar{\mathcal P}^{\bar\mu\bar\nu}\equiv 
\frac{\bar k^{\bar\mu}\bar k^{\bar\nu}}{\bar k_T^2}.
\ee
Eq.~\href{coeff_hadro} should be compared to its photo- or 
lepto-production counterpart
Eq.~\href{coeff}. In particular, we now have (see Eq.~\href{onshell})
\be\label{onshell_hadro}
\lim_{\xi, \bar\xi \rightarrow 0} 
\left<C\lp\frac{x}{z\zbar}, \xi, \bar \xi,\as \rp\right>_{\theta,\bar\theta} =
\sigma_{\text{on-shell}}\lp g(zn), g(zp) \rightarrow \mathcal F \rp.
\ee

Starting from the hard coefficient function Eq.~\href{coeff_hadro}, we can repeat
the same steps of the previous subsections, and arrive at  the hadroproduction
formula (with $\mu^2=Q^2$)
\bea\label{l_iter_hadro}
&&\sigma\lp N, \as(Q^2) \rp = 
\int_0^\infty d\xi R\lp\frac {\as(Q^2)} N \rp\gamma\lp 
\frac {\as(Q^2)} N \rp\xi^{\gamma\lp \frac {\as(Q^2)} N \rp -1} \times
\nonumber\\
&&\quad\times
\int_0^\infty d\bar\xi R\lp\frac {\as(Q^2)} N \rp\gamma\lp 
\frac {\as(Q^2)} N \rp\bar\xi^{\gamma\lp \frac {\as(Q^2)} N \rp -1} 
C\lp N, \xi, \bar \xi, \as(Q^2)\rp ,
\eea
to be compared to its photo- or lepto-production counterpart Eq.~\href{l_iter}.

Before concluding this section, we now briefly account for quark-initiated 
processes. The analysis in~\cite{BFKL} shows that at the LL$x$ level only gluons
enter in the generalized ladder, with one exception: since the $P^0_{gq}$ 
splitting function is singular at small $x$, the first rung of the ladder
can be a quark. In other words, the initial quark can emit a collinear quark,
thus becoming a gluon which starts radiating the generalized ladder 
Fig.~\ref{genlad_pic}, leading to a cross-section proportional to $P^0_{gq}$. 
Since at small $x$ we have $P^0_{gq}= \frac{C_F}{C_A} 
P^0_{gg}$, the only difference between quark- and gluon-initiated processes
is a color-charge factor.

\section{ High energy factorization of  rapidity distributions}
\label{rapidity}

The formulation of high energy factorization of Sect.~\ref{inclusive}
will now be extended to rapidity distributions, thereby leading to a
method for the determination of LL$x$ contributions to all orders.
The argument will closely follow that of 
Sect.~\ref{inclusive}.

\subsection{Kinematics and factorization}

We now concentrate from the onset on hadroproduction of
a final state system $\mathcal{S}$ 
with invariant mass $Q^2$:
\be
h_1(P_1) + h_2(P_2) \to \mathcal{S}(p_\mathcal{S})+X\,
\label{hadroproduction}
\ee
and we study distribution in the rapidity of $\mathcal{S}$
\be\label{rapidity_def}
Y \equiv \frac{1}{2} \ln \frac{E_\mathcal{S}+p_{\mathcal{S}z}}{E_\mathcal{S}-p_{\mathcal{S}z}}\,,
\ee
where $E_{\mathcal S}$, $p_{\mathcal{S} z}$ are measured in the hadronic 
center-of-mass frame. 
The (dimensionless) 
hadronic rapidity distribution is written in terms of parton
distributions $f_i^h$ and a partonic rapidity distribution  $\frac{\d
  \sigma_{ab}}{\d y}$: 
\bea\label{pdfconv}
&&\frac{\d \sigma_h}{\d Y}(x_h,Y,Q^2,\as) = 
\sum_{a,b} \int_{\sqrt{x}_h e^Y}^{1} d x_1 \int_{\sqrt{x}_h e^{-Y}}^{1} d x_2 
\times\nonumber\\
&&\quad\times
\frac{\d \sigma_{ab}}{\d y}
\lp \frac{x_h}{x_1 x_2}, \frac{\mu^2}{Q^2},Y-\frac{1}{2} \ln\frac{x_1}{x_2},
\as\rp \times f^{h_1}_a(x_1,\mu^2) f^{h_2}_b(x_2,\mu^2)\,.
\eea
where $x_h= \frac{Q^2}{S}$ and $S=(P_1+P_2)^2$. Note that we have used
the well-known fact that in collinear factorization the hadronic
and partonic center-of-mass frames are related by a longitudinal boost.

Henceforth, as in the previous section, we work in the partonic
center-of-mass frame, where $x_1=x_2=1$ (thus the partonic rapidity $y$
is equal to the hadronic one $Y$) and  
the partonic rapidity distribution is given by 
\be\label{rap_def}
\frac{\d \sigma}{\d y} \lp x, \frac{\mu^2}{Q^2}, y, \as\rp
\equiv \int \frac{\overline{|M|^2}}{2s}  d\Pi_f 
\delta \lp y-
\frac 1 2 \ln \frac {E_{\mathcal{S}}+p_{\mathcal S z}}{E_{\mathcal{S}}-p_{\mathcal S z}}\rp
\ee
where the bar stands for sum (average) over final (initial) helicities
and colors and $d \Pi_f$ is the final particle phase space, and here
and henceforth for simplicity the parton indices
 $a,b$ have been omitted.

Resummation at the differential level in rapidity can now be performed
by repeating the argument of Sect.~\ref{inclusive}, but with the
partonic cross-section now multiplied by the kinematic constraint
enforced by the rapidity delta in Eq.~(\ref{rap_def}). In
the factorized expression Eq.~\href{fact2GI_hadro} the 
dependence on the momentum $p_{\mathcal S}$ of the final
state system $\mathcal S$ (see Fig.~\ref{hard_pic})
is entirely contained in the hard
part. Therefore, the kinematic constraint only enters the definition
of the coefficient function $C$ which now becomes
\bea
&&C_y\lp x, z, \bar z, \xi, \bar \xi, y, \as \rp
=\int_0^{2\pi} \frac{\d \theta}{2\pi} 
\int_0^{2\pi}\frac{d\bar \theta}{2\pi}
\frac{Q^2}{2 s z \zbar } \times\nonumber\\
&&\quad\times\left[
\mathcal P_{\mu\nu} \bar {\mathcal P}_{\bar\mu\bar\nu} 
H^{\mu\nu\bar\mu\bar\nu}(n_L, p_L, p_{\mathcal F}, \as)\right]
\delta \lp y-
\frac 1 2 \ln \frac {E_{\mathcal S}+p_{\mathcal S z}}{E_{\mathcal S}-p_{\mathcal S z}}\rp,
\label{cdiffdef}
\eea
where $z$, $\bar z$, $\xi$, $\bar \xi$ were respectively defined in
Eqs.~(\ref{Sudakov}, \ref{Sudakov_n}, \ref{x_xi_defs}, \ref{xi_proj_hadro})
 and we have made explicit the dependence on the rapidity $y$,
which in turn is fixed in terms of the remaining kinematic variables
by the delta in Eq.~(\ref{rap_def}). Note that, as in the inclusive
case   Eq.~\href{coeff_hadro}, in general the coefficient 
function $C_y$ also depends on a residual set
 $\Omega_{\mathcal F}$ of final-state variables which survive the phase space
integration. This dependence is  omitted for simplicity.

Equations~(\ref{coeff_hadro}, \ref{cdiffdef}) show that the ladder part
is unaffected by the rapidity constraint. The rapidity-dependent
coefficient function $C_y$ instead
acquires a nontrivial dependence on all kinematic variables because of
this constraint. However, we will now show that invariance under
longitudinal boosts of the ladder 
implies that the dependence of $C_y$ on the
longitudinal momentum fractions $z$, $\bar z$ is completely determined
by its dependence on the rapidity $y$ and the scaling variable
$x$.  
But the argument presented in Sect.~\ref{inclusive} shows that 
the effect of the ladder insertion on the external legs (which
generates the LL$x$ series) can be cast
in terms of  QCD evolution in $k_T^2$, characterized by standard
anomalous dimensions. Therefore, this implies that we will be able to 
trade the rescaling of the
momenta $p_L$ and $n_L$ which enter the hard part for a shift in
rapidity. This will enable us to determine fully the LL$x$ resummation
of the rapidity distribution from the knowledge of the rapidity
dependence of the hard part $C_y$.

In the remainder of this section, we will prove these results, which
are essentially of kinematic origin, while
the subtraction of collinear singularities in the ladder part is
unchanged and can be handled as in Sect.~\ref{inclusive}.
Therefore, for clarity we will simply regulate collinear singularities
with a cutoff. The price to pay for this is that we will loose full
control of the factorization scheme. The full computation in the
$\msb$ scheme is presented in Appendix~\ref{app_msbrap}, and will
provides us with a determination of the scheme-dependent factor
$R$.

\subsection{Ladder part and rapidity dependence}
\label{llxrap}

As we have seen in Sect.~\ref{inclusive}, large logs of
$\frac{1}{x}$ come from gluon ladders attached to the external legs
  of the hard part. We now show how to compute the effect of these
  ladders on rapidity distributions. The structure of the computation
  is unchanged: the ladder is constructed by iterating $n$ times a
  kernel $K$, which is closely related to the standard QCD anomalous
  dimension.   In the case
 of single insertion of the kernel  on the lower leg (say) 
 we have 
\bea\label{rap_NLO}
&&\frac{\d\sigma^1}{\d y}\lp x, \frac{\mu^2}{Q^2}, y, \as \rp
=\nonumber \\
&&=\quad
\int_x^1 \frac{dz}{z} \int_{\mu^2}^\infty \frac{dk_T^2}{k_T^2}
C_y\lp \frac{x}{z}, \frac{k_T^2}{Q^2},
y\lp x, z, \frac{k_T^2}{Q^2} \rp, \as\rp K\lp z,  \as\rp, 
\eea
which only differs from Eq.~\href{1gluon_xspace} of the inclusive case
because of the replacement of the inclusive coefficient function $C$
with the differential one $C_y$, and because of the kinematic
constraint imposed by the delta function Eq.~(\ref{rap_def})
which relates the rapidity $y$ to the
other kinematic variables of the hard part.

\begin{figure}
\centering
\epsfig{file=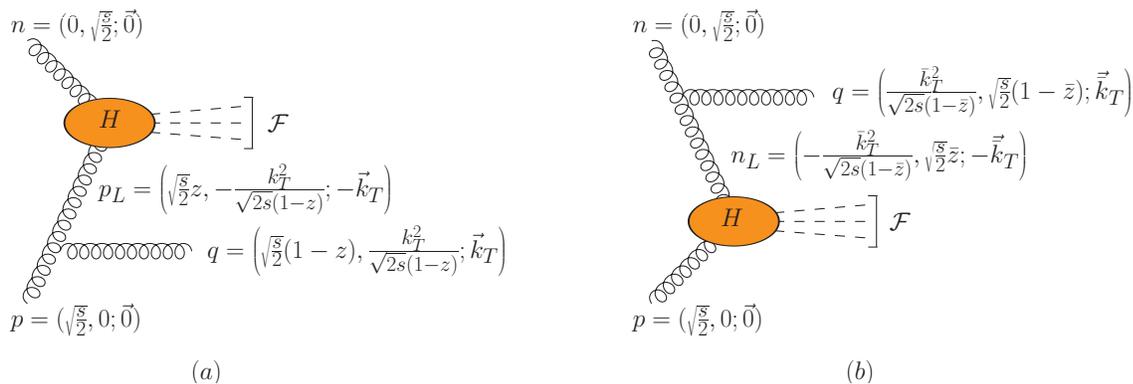,width=1.0\textwidth}
\caption{Single emission kinematics from both legs in terms of light-cone
Sudakov components, see Eq.~\href{Sudakov}.}
\label{NLO_pic}
\end{figure} 

The kinematics of the single insertion is displayed in
Fig.~\ref{NLO_pic}, using the notation of Eq.~(\ref{Sudakov}), 
in the LLQ$^2$ case in which the kernel corresponds to
single-gluon emission as discussed in Sect.~\ref{genladdl}. As
discussed in Sect.~\ref{llxladder} the kinematics in the general LL$x$
case is the same. It is apparent that 
\be
p_L=\lp\sqrt{\frac{s}{2}}
  z,0;-\vec k_T\rp \lp1+O\lp\frac{k_T^2}{s}\rp\rp. 
\ee
It follows that the longitudinal components of $p_L$  are related to
those of the Sudakov base momentum $p$ by a longitudinal
boost. Therefore
\be\label{rap_long_boost}
C_y(x, z, k_T^2, Q^2,y\lp x, z, \frac{k_T^2}{Q^2} \rp,\as) = 
C_y\lp \frac{x}{z},\frac{k_T^2}{Q^2}, y\lp x, \frac{k_T^2}{Q^2} \rp- 
\frac 1 2 \ln z, \as\rp
+ O\lp \frac{k_T^2}{s} \rp,
\ee
i.e.,  the effect of a $K$-kernel insertion on the rapidity
dependence of the 
hard part is 
a longitudinal boost, up to subleading terms in small $x$ kinematics
Eq.~(\ref{sxkin}). 

Eq.~\href{rap_NLO} can be easily generalized to the case of $n$-kernel 
insertion, with the result:
\bea\label{1leg_fac}
&&\frac{\d\sigma^{n}}{\d y}\lp x, \frac{\mu^2}{Q^2}, y, \as \rp
=\int_{\mu^2}^\infty \frac{dk_T^2}{k_T^2}\frac{1}
{(n-1)!}\ln^{n-1}\frac{k_T^2}{\mu^2}\times\nonumber\\
&&\quad \times 
\int_{x}^1\frac{dz_n}{z_n} P\lp z_n, \as\rp
\times...\times
\int_{\frac{x}{z_2...z_n}}\frac{dz_1}{z_1} P\lp z_1, \as\rp
\times\nonumber\\
&&\quad \times C_y\lp \frac{x}{z_1 ...z_n},\frac{k_T^2}{Q^2}, 
y -\frac 1 2 \ln z_1 - ... -
\frac 1 2 \ln z_n , \as\rp,
\eea
where $P$ is the inverse Mellin transform of the kernel $K$ and 
for simplicity we have written $y(x, k_T^2/Q^2) = y$.

Equation~\href{1leg_fac} can be factorized by taking 
a Fourier-Mellin transform, defined (together with  its inverse) as
\bea\label{f-m}
f(N,b)&=& \int_0^1 dx x^{N-1}\int_{-\infty}^\infty dy e^{i b y}f(x,y);\nonumber\\
f(x,y)&=&\int_{c-i \infty}^{c+i\infty} \frac{dN}{2\pi i} x^{-N}
\int_{-\infty}^{\infty}\frac{db}{2\pi} e^{-i b y}f(N,b),
\eea
where as usual by slight abuse of notation we use the same symbol to
denote the function and its transform.
We get
\bea\label{1leg_FMtransf}
\frac{\d\sigma^n}{\d y} \lp N, \frac{\mu^2}{Q^2}, b, \as\rp
&=& \int_{-\infty}^\infty dy ~ e^{i b y}  \int_0^1 dx ~ x^{N-1}
\frac{\d \sigma^n}{\d y} \lp x, \frac{\mu^2}{k_T^2}, y, \as \rp=
\nonumber\\
&=&  \int_{-\infty}^\infty dy ~ e^{i b y} 
\int_0^1 dx ~ x^{N-1}\times
\int_x^1 \frac{dz_n}{z_n} e^{i \frac {b} 2 \ln z_n}
P\lp z_n, \as\rp\times
...\times\nonumber\\
&&\qquad\times
\int_{\frac{x}{z_2...z_n}}\frac{dz_1}{z_1}e^{i\frac b 2 \ln z_1}
P\lp z_1, \as, \rp \times \int_{\mu^2}^\infty \frac{dk_T^2}{k_T^2}
\frac{\ln^{n-1}\frac{k_T^2}{\mu^2}}{(n-1)!}\times
\nonumber\\
&&\qquad\quad \times
C_y\lp \frac{x}{z_1 ...z_n},\frac{k_T^2}{Q^2}, 
y,\as\rp.
\eea
Each $z$ integral in Eq.~\href{1leg_FMtransf} is a convolution, hence in
Fourier-Mellin space the result factorizes:
\be\label{1leg_FM_ngluon}
\frac{\d\sigma^n}{\d y} 
\lp N, \frac{\mu^2}{Q^2},b,\as \rp = 
\left[\gamma \lp N+\frac{ i b}{2}, \as \rp\right]^n
\int_{\mu^2}^{\infty}\frac{dk_T^2}{k_T^2}
C_y\lp N, \frac{k_T^2}{Q^2}, b,\as\rp 
\frac {\ln^{n-1} \frac{k_T^2}{\mu^2}}{(n-1)!}.
\ee

As in Sect.~\ref{inclusive}, the final result for the cross-section is
obtained by adding up the contributions from $n$ insertions of the
kernel, with $1\le n\le\infty$. We get
\bea\label{1leg_FM}
&&\frac{\d \sigma}{\d y}
\lp N, \frac{\mu^2}{Q^2}, b,\as \rp 
 = \sum_{n=1}^{\infty} \frac{\d \sigma^n}{\d y} = \nonumber\\
&&\quad
= 
\gamma\lp \frac{\as}{N+\frac{ i b}{2}}\rp
\int_{\mu^2}^\infty dk_T^2 (k_T^2)^{\gamma\lp \frac{\as}{N+i b/2} \rp -1}
C_y\lp  N,\frac{k_T^2}{Q^2}, b,\as\rp.
\eea

\subsection{Resummation of rapidity distributions}
\label {resrapd}

Equation~(\ref{1leg_FM}) provides the resummation of LL$x$ terms
due to radiation from one of the two outer legs of the hard kernel, so
a full resummation can be obtained from its
two-leg generalization. Before
discussing this, we observe that in Eq.~(\ref{1leg_FM}) the multiple
poles in $N=0$ which in the inclusive result Eq.~(\ref{finresgen}) lead
to the LL$x$ terms have now become  multiple poles in $N=-i b/2$.
On top of these, the hard coefficient function $C_y$ may still contain
poles in $N=0$, but, as discussed in Sect.~\ref{sec_hard_part}, this only
happens in the somewhat pathological case of pointlike interactions.

In order to understand how poles in $N=-i b/2$ can be related to large
logs of $x$ we note that 
the value of $x$ determines the range of the partonic rapidity:
\be\label{raprange}
\frac{1}{2} \ln x\le y \le\frac{1}{2}\ln\frac{1}{x}.
\ee
It follows that as the energy increases, i.e. as $x$ 
decreases, the rapidity range widens. 
This implies that high energy enhanced terms are not necessarily
logarithmic at the level of rapidity distributions: for
instance, a flat rapidity distribution leads to a contribution enhanced by
 $\ln x$  in the inclusive cross-section because of this
widening of the phase space.  
This can also be seen in Fourier-Mellin space, by noting that (see
Eq.~\href{f-m})  the Mellin-space
inclusive cross-section is found letting $b=0$ in the 
Fourier-Mellin space rapidity distribution.
Therefore poles at $N=0$ in the inclusive result can be obtained from
any  term of the form
$\lp \frac \as {N+f(b)}\rp^k$ such that $f(0)=0$.

We now turn to terms coming from emissions from the other leg.
These are easily determined: the only difference 
in comparison  to Eq.~\href{rap_NLO} is that the boost has now
parameter $1/\bar z$ 
(see Fig.~\ref{NLO_pic}).
The result is  then  identical to Eq.~\href{1leg_FM}, but with 
the replacement:
\be
\gamma\lp \frac{\as}{N+ib/2} \rp \longrightarrow 
\gamma\lp \frac{\as}{N-ib/2} \rp. 
\ee

The resummation prescription is thus the following.
First, the rapidity distribution for the process
$g^*(p_L)+g^*(n_L)\rightarrow \mathcal S$ is determined 
with off-shell incoming gluons. 
The rules for computing $C_y$ are the same as presented in
Sect.~\ref{sec_hard_part} 
for the inclusive case, namely, the incoming  
off shell momenta  are $p_L = p + k,~p_L^2 = -k_{T}^2$ and
$n_L = n + \bar k,~n_L^2 = -\bar k_{T}^2$, and the 
average over the initial gluon polarizations is performed by the projectors 
$\mathcal P_{\mu\nu}$ Eq.~(\ref{proj}) and 
$\bar {\mathcal P}_{\bar \mu \bar \nu}$ Eq.~\href{xi_proj_hadro}. 
The coefficient function $C_y$
is then defined using Eq.~(\ref{cdiffdef}).

In terms of the differential coefficient function $C_y$ the resummed
rapidity distribution is given by the two leg generalization of 
Eq.~\href{1leg_FM}:
\bea\label{2legrap}
&&\frac{\d\sigma}{\d y}
\lp N, \frac{\mu^2}{Q^2}, b,\as\rp
 =
\int_{\mu^2}^\infty d k_T^2
\gamma\lp \frac{\as}{N+ \frac{i b}{2}} \rp 
(k_T^2)^{\gamma\lp \frac{\as}{N+ i b/2} \rp -1} \times\nonumber \\
&&\quad \times\int_{\mu^2}^\infty d \bar k_T^2
\gamma \lp \frac{\as}{N- \frac{i b}{2}} \rp 
(\bar k_T^2)^{\gamma\lp \frac{\as}{N- i b/2} \rp -1} 
C_y\lp N,\frac{k_T^2}{Q^2},\frac{\bar k_T^2}{Q^2}, b, \as\rp.
\eea
This result has been obtained with a collinear cutoff $\mu^2$. 
Although the full
$\msb$ derivation is given in Appendix~\ref{app_msbrap}, here we give
the final result:
\bea\label{2legrap_msb}
&&\frac{\d\sigma_{\msb}}{\d y}
\lp N,  b,\as\rp = 
\int_0^\infty d \xi
\gamma\lp \frac{\as}{N+ \frac{i b}{2}} \rp 
R_{\msb}\lp \frac{\as}{N+\frac{i b}{2}} \rp 
\xi^{\gamma\lp \frac{\as}{N+ i b/2} \rp -1} \times\nonumber \\
&&\quad \times\int_0^\infty d \bar \xi
\gamma \lp \frac{\as}{N- \frac{i b}{2}} \rp 
R_{\msb}\lp \frac{\as}{N-\frac{i b}{2}} \rp 
\bar \xi^{\gamma\lp \frac{\as}{N- i b/2} \rp -1} 
C_y\lp N,\xi,\bar\xi, b, \as\rp,\label{rap_res_msb}
\eea
where $\xi=k_T^2/Q^2$, $\bar \xi= \bar k_T^2/Q^2$ and we have chosen
$\as=\as(Q^2)$ and consequently dropped the explicit 
$\mu^2/Q^2$ dependence.

We
 can cast Eq.~\href{2legrap_msb} as an $M$-Mellin transform of the off-shell
rapidity distribution $C_y(N,\xi,\bar \xi,b,\as)$ in a form similar to that
Eqs.~(\ref{impfac}, \ref{mfixing}) of the inclusive case, by defining
a double $M$-Mellin transform of the differential coefficient function $C_y$.
Setting again $\mu^2=Q^2$ we get
\beq
h_y(N,M,\bar M,b,\as) \equiv M  \bar M 
\int_0^\infty d\xi
\xi^{M -1} 
\int_0^\infty d\bar\xi
\bar\xi^{\bar M -1} C_y(N,\xi,\bar\xi, b,\as) R(M)R(\bar M),
\label{impfacdiff}
\eeq
with again $\as=\as(Q^2)$. 
The resummed result is then found from Eq.~(\ref{impfacdiff})
through the identifications
\bea
M &=& \gamma\lp \frac{\as}{N + \frac{i b}{2}}\rp\nonumber\\
\bar M &=& \gamma\lp \frac{\as}{N - \frac{i b}{2}}\rp\label{m12fixing}, 
\eea
namely
\be
\frac{d\sigma}{dy} \lp N,b,\as(Q^2)\rp
= h_y\left(N,\gamma\lp \frac{\as(Q^2)}{N + \frac{i b}{2}}\rp,
\gamma\lp \frac{\as(Q^2)}{N -\frac{i b}{2}}\rp,b,\as(Q^2)\right) .
\label{catanilikeresum}
\ee
The inclusive result (i.e. the two leg generalization of 
Eq.~(\ref{cataniresum})) is recovered from the
differential one Eq.~(\ref{catanilikeresum}) by setting $b=0$.
Resummed results at the hadronic level are finally obtained by
substituting the resummed differential coefficient function
Eq.~(\ref{catanilikeresum}) in the factorization formula
Eq.~(\ref{pdfconv}).

As in the inclusive case, the factorized expression
Eq.~(\ref{rap_res_msb}) can be viewed as a factorization theorem, by
rewriting it as
\be\label{rap_res_Catani}
\frac{\d \sigma}{\d y} = \int\frac{dz}{z} \frac{d k_T^2}{k_T^2}
\int \frac{d\bar z}{\bar z}\frac{d \bar k_T^2}{\bar k_T^2}
C_y\lp \frac{x}{z \bar z},\frac{k_T^2}{Q^2}, \frac{\bar k_T^2}{Q^2},
 y - \frac{1}{2} \ln \frac {z}{\bar z} \rp
\mathcal G\lp z, k_T^2\rp \mathcal G\lp \bar z, \bar k_T^2\rp.
\ee
However, a direct derivation of Eq.~(\ref{rap_res_Catani}) from the
determination of the gluon Green function and iteration of the BFKL
kernel from its inclusive counterpart appears to be less immediate
than our derivation. This is because in our case
the rapidity dependence along the ladder, which has the standard $k_T$
ordered structure, is trivial, while the rapidity flow along a BFKL
ladder is less obvious.

\section{Higgs rapidity distribution at small $x$}
\label{sec_higgs_res}

As a first application of the formalism developed in the previous
part of the paper, we will consider Higgs production in  gluon-gluon
fusion. 
This process is particularly simple kinematically 
because the final state $\mathcal F$ contains just  one particle, 
the Higgs. In contrast, recalling  from Sec.~\ref{inclusive}
that $\mathcal F$ is the
lowest-order final state for the given inclusive process with at least
one incoming gluon, it is clear that for Drell-Yan~\cite{SimoneDY} or
prompt photon~\cite{GiovanniPhoton} production 
$\mathcal F$  is a two-particle 
final state. 

\begin{figure}
\centering
\epsfig{file=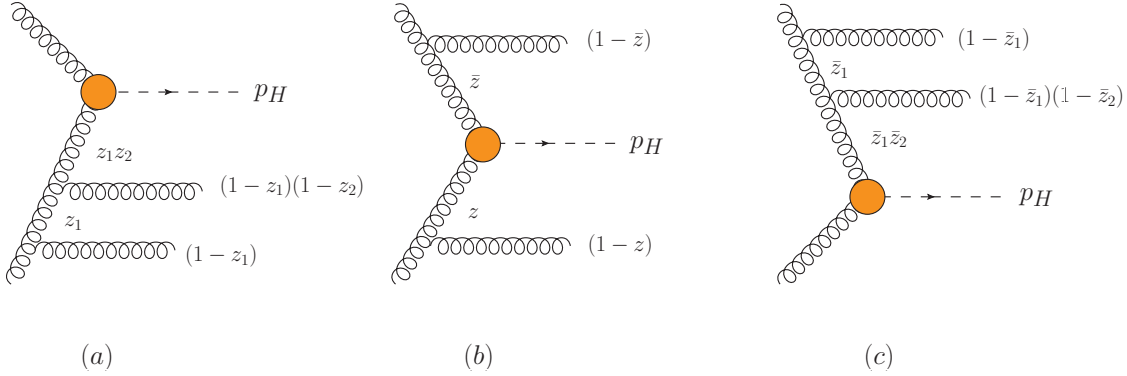,width=1.0\textwidth}
\caption{Graphs contributing to the small $x$ Higgs rapidity distribution 
at NNLO. Here the (orange) blob stands for a $H G^a_{\mu\nu}G^{a,\mu\nu}$ 
effective coupling (heavy top limit) or for the top quark loop 
(full Standard Model with finite $m_{\rm top}$).
Only relevant light-cone fractions (in units of $\sqrt{s/2}$) are reported.}
\label{NNLO_pic}
\end{figure}

In the Standard Model the Higgs couples to gluons 
via a heavy quark loop. If the top mass $m_{\rm top}$ is much bigger than the 
Higgs mass one can describe the interaction through the
effective Lagrangian
\be\label{HEFT}
\mathcal L_{int}=\frac{\sqrt{G_F\sqrt{2}}\as}{12\pi}
H G^a_{\mu\nu} G^{\mu\nu a} 
\ee
which corresponds to a pointlike interaction between the gluons and 
the Higgs. 
As it is well-known~\cite{KramerHeavyTop} this approximation is reliable
within 10\% for Higgs masses below 1~TeV. However, at small $x$ this 
approximation fails: when the partonic center-of-mass energy squared $s$ is
large enough (i.e. at small $x$) gluons can resolve the quark loop. 
Consequently, the effective theory at small $x$ leads to spurious double
logarithmic singularities~\cite{CataniHQ,HautmannHiggs}, 
while the full theory has single
log behavior:
\be\label{pl_res}
\sigma \sim \sigma_0\times\left\{
\begin{array}{cc}
 \ \delta(1-x) + \sum_k c_k \as^k \ln^{2k-1}\frac 1 x
& m_{\rm top}\rightarrow \infty\\
 \delta(1-x) + \sum_k d_k \as^k \ln^{k-1}\frac 1 x 
& \text{finite }m_{\rm top} 
\end{array}
\right.
\ee

The treatment of rapidity distributions for this case is particularly
simple due to the fact that,
using the small $x$ kinematics Eq.~\href{sxkin},
the rapidity is entirely fixed in terms
of the momentum fractions of incoming partons.
Indeed, for single emission (see Fig.~\ref{NLO_pic}) we have
\be\label{y_up}
y = \frac{1}{2} \ln \frac{p_H^+}{p_H^-}= 
\frac 1 2 \ln \frac{1-\frac{\bar k_T^2}{s(1-\bar z)}}{\zbar} =
\frac 1 2 \ln \frac 1 \zbar + O \lp\frac{\bar{k}_T^2}{s}\rp + O \lp\zbar\rp 
\ee
if the emission is from the upper leg and
\be\label{y_down}
y = \frac 1 2 \ln \frac{z}{1-\frac{k_T^2}{s(1-z)}} =
\frac 1 2 \ln z + + O \lp\frac{k_T^2}{s}\rp + O \lp z\rp
\ee
if the emission is from the lower leg. 
In case of a single emission from each of the incoming
legs (see Fig.~\ref{NNLO_pic}-b), we have
\be\label{y_twolegs_full}
y = \frac 1 2 \ln \frac{z - \frac{\bar k_{T}^2}{s(1-\zbar)}}
{\zbar + \frac{k_{T}^2}{s(1-z)}} = 
\frac 1 2 \ln \frac z \zbar + \frac 1 2 
\ln\frac{1-\bar \xi\frac{x}{z\zbar} \zbar}
{1-\xi\frac{x}{z\zbar} z},
\ee
with $Q^2=m_H^2$ (hence $x=m_H^2/s$) for Higgs production. But using 
momentum conservation 
\be
\frac{x}{z\zbar} = \frac{m_H^2}
{m_H^2 + \lp \mathbf {k}_{T}+ \mathbf {\bar k}_{T} \rp^2} = 
\frac 1 {1 + \xi + \bar \xi + 2 \sqrt{\xi \bar \xi}\cos\theta}
\ee
so  $\xi x/(z\zbar)$, $\bar\xi x/(z\zbar)$ is at most $O(1)$ and therefore
\be\label{y_twolegs}
y = \frac 1 2 \ln \frac z \zbar + O(z,\zbar).
\ee
Equations~(\ref{y_up}-\ref{y_twolegs}) show that for Higgs
production the cross-section at fixed $y$ is just a cross-section at fixed 
$z$, $\zbar$.
Of course, these purely kinematic relations hold for
any $2\rightarrow 1$ process.

In this section, we compute the small $x$ behavior of the Higgs
cross-section (in the $gg$ channel) 
up to NNLO, before turning to the all-order resummation
in the next section. In the $m_{\rm top}\to\infty$ limit, 
we will obtain analytic closed-form results which at NLO may be
compared to the known~\cite{AnastasiouHiggsNLO} full analytic
expression. For finite $m_{\rm top}$ we will reduce the NLO and NNLO
small $x$ coefficients to quadratures and tabulate them numerically as
a function of $\tau$ (i.e. of the Higgs mass, see Eq.~\href{def_tau}). In this case a
direct comparison to the known NLO result~\cite{AnastasiouHQ} 
is not possible
because the latter is only known in the form of the numerical integral
of a fully differential cross-section.

\subsection{Heavy top approximation}
\label{mtopinfty}

In the infinite $m_{\rm top}$ approximation the NLO
Higgs rapidity distribution is known in closed analytic
form~\cite{AnastasiouHiggsNLO}.  Here we 
reproduce the NLO result in the small $x$ limit. As we have mentioned
in Sec.~\ref{inclusive}, only ladder-type diagrams contribute to LL$x$. 
Hence we must  compute the two diagrams Fig.~\ref{NLO_pic}.
We start with
emission from the upper leg. Up to power-suppressed terms in  $\bar z$ and
$\bar k_T^2/s$, the fully differential cross-section is
\bea\label{higgs_NLO}
d\bar\sigma_\text{1-leg} &=& \left[ \frac{\sigma_0 \zbar^2 s^2}
{2 s \zbar} 
\times \frac{2\pi}{m_H^2}\delta \lp 1-\frac{\zbar}{x}+
\frac{\bar k_T^2}{m_H^2}\rp
\right] \times
\left[
\frac{\as \mu^{2\epsilon}N_c}{\pi}\frac{d\zbar}{\zbar} 
\frac{d \bar k_T^2}{\lp \bar k_T^2\rp^{1+\epsilon}}\frac{(4\pi)^\epsilon}{\Gamma(1-\epsilon)}
\right]=\nonumber\\
&=&
\left[\sigma_0\frac \zbar x 
\delta \lp 1 - \frac \zbar x + \bar \xi \rp\right]
\left[\asb  \frac {d\zbar}{\zbar} \frac{d\bar\xi}{\bar\xi^{1+\epsilon}} 
\frac{(4\pi)^\epsilon}{\Gamma(1-\epsilon)}\right],
\eea
where we have defined 
\be
\label{bornheft}
\sigma_0\equiv G_F\sqrt{2} \as^2/576 \pi
\ee
and in the last step we have set $\mu^2=Q^2=m_H^2$ for simplicity.
As in Sec.~\ref{inclusive},  the bar over $\sigma$ indicates 
that it contains an unsubtracted collinear singularities, so
Eq.~\href{higgs_NLO} only makes sense at a regularized level.

Note that the hard part, contained in
the first square bracket in
Eq.~\href{higgs_NLO}, reduces to the LO result when $\bar k_T^2=0$,
$\zbar=1$, as it ought to according to 
Eq.~\href{onshell}. The only $\bar k_T$ 
dependence of the hard part comes from the phase space
delta function, i.e. the hard amplitude is $\bar k_T$ independent. This makes
high energy factorization almost trivial in this case.
Indeed, to compute the rapidity distribution, we just have to insert in the
differential cross-section Eq.~\href{higgs_NLO} the rapidity $\delta$ 
function (see Eq.~\href{rap_def}), which in our case is
\bea
\delta\lp y - \frac 1 2 \ln \frac{p_H^+}{p_H^-} \rp &=& \delta \lp y - \frac 1 2 \ln \frac 1 \zbar \rp\nonumber\\
&=& 2 u \delta \lp \zbar - u \rp,
\label{rapdeltah}
\eea
where, following Ref.~\cite{AnastasiouHiggsNLO}~\footnote{Note that even though we
follow closely the notation of Ref.~\cite{AnastasiouHiggsNLO}, we
depart from it in  calling $y$  the rapidity, whereas in
Ref~\cite{AnastasiouHiggsNLO} $y$ denotes a certain function of $u$
and $z$.}, we have defined 
\be
u\equiv\exp (-2 y)
\label{udef}
\ee
and the first equality in Eq.~\href{rapdeltah} holds  up to subleading
terms.
In terms of $u$ the kinematic limits Eq.~(\ref{raprange}) on the
rapidity become
\be
x \le u \le \frac 1 x.
\label{raprangeu}
\ee
The transformation $y\leftrightarrow -y$  corresponds to $u\leftrightarrow 1/u$ 
symmetry. For the emission from the upper leg one obtains the 
$x\le u \le 1$ part of the rapidity distribution (as it can be immediately seen 
from the equality $u=\zbar$), while emission from the lower leg covers
the other half of the rapidity range. 

Using the rapidity 
$\delta$ function to perform the $\zbar$ integration and the momentum
$\delta$ function to perform the $\bar \xi$ integration we 
immediately obtain the Higgs rapidity distribution as a function of $u$:
\be\label{oneleg_du}
\frac{\d\bar\sigma_\text{1-leg}}{\d u}(x,u) = \frac{1}{2u} \frac{\d\sigma}{\d y} = 
\sigma_0 \asb \frac{(4\pi)^\epsilon}{\Gamma(1-\epsilon)}
\frac{x^\epsilon}{(u-x)^{1+\epsilon}},
\ee
where  we have now omitted the explicit dependence 
on $\as$ for simplicity, as well as the dependence on $\mu^2/Q^2$ because 
we are working in the case $\mu^2=Q^2=m_H^2$. 
For $\epsilon=0$ Eq.~\href{oneleg_du} has an endpoint rapidity 
singularity in $u=x$. 
We regularize it by defining the plus-prescription
\bea
\int_x^1 du \frac{f(u)}{(u-x)^{1+\epsilon}} &=&
\int_x^1 du \frac {f(u)-f(x)}{(u-x)^{1+\epsilon}} 
+ f(x)\int_x^1 du \frac{1}{(u-x)^{1+\epsilon}} \equiv \nonumber \\
&\equiv& \int_x^1 du f(x) \left[
\frac{1}{(u-x)_+} - \frac{1}{\epsilon} \delta(x-u)(1-x)^\epsilon
\right],
\label{upluspresc}
\eea
where the distribution is in $u$ space. Using Eq.~(\ref{upluspresc})
and performing the $\msb$ subtraction our result reduces to
\be\label{du}
\frac{\d\sigma_\text{1-leg}}{\d u}(x,u) = \sigma_0 \asb \left[
\frac{1}{(u-x)_+} - \delta(u-x) \ln x
\right]
\ee
up to terms which vanish like powers of $x$. In Eq.~\href{du} we have taken
the limit $\epsilon\to0$ (after $\msb$ subtraction). 
Integrating Eq.~\href{du} over $u$, the first term 
gives a vanishing contribution, while the delta function 
integration leads to $\sigma = \sigma_0 \asb \ln 1/x$, in agreement 
with the inclusive result~\cite{HautmannHiggs}. 

One  may be tempted
to think that in the small $x$ limit
the first term on the right-hand side of Eq.~(\ref{du}) can be
neglected in comparison to the second one. However,
 this is incorrect because the first term
is phase space enhanced. To see this, we use
Fourier-Mellin space. Start with the first term on
the right-hand side of Eq.~(\ref{du}). Its
Fourier transform with respect
to  $u$ is
\bea\label{four_first}
&&\int u^{-\frac {ib} 2}\left[\frac{1}{(u-x)_+}\right] = 
\frac{2 i}{b} - \lp \gamma_E + i \pi + \frac{2\pi i}{e^{b\pi}-1}\rp 
x^{- \frac {ib} 2} + \nonumber\\
&&\quad 
- x \frac{_2F_1\lp 1,1+\frac{ib} 2, 2 + \frac{ib} 2, x\rp}{1 + \frac{ib} 2}
- x^{-\frac{ib} 2}\ln\frac{1}{x}-x^{-\frac{ib} 2} \ln(1-x) 
- x^{-\frac{ib} 2}\psi\lp 1 - \frac{ib} 2 \rp .\nonumber\\
\eea
Note in particular that the $\ln 1/x$ term in the last line exactly cancels
the Fourier transform of the delta $\delta(u-x) \ln 1/x$ contribution 
in Eq.~\href{du}. 
Performing the Mellin transform, we get 
\be
\mathcal M\mathcal F \left[ \frac 1 {(u-x)_+} \right] = 
\frac{1 - N H(N) + N H\lp N+ \frac{ib} 2 \rp}{N\lp N-\frac{ib} 2\rp}
- \frac{1}{\lp N - \frac{ib} 2 \rp^2},
\ee
where $H(N)$ is the $N$-th harmonic number.
Turning now to the second term on
the right-hand side of Eq.~(\ref{du}), its 
Fourier-Mellin transform is
\be\label{mf_delta}
\mathcal M\mathcal F \left[
\delta(u-x) \ln \frac 1 x
\right] =
\frac{1}{\lp N - \frac{i b }{2} \rp^2}.
\ee
Combining the two terms we get
\bea
&&\mathcal M \mathcal F \left[ \frac 1 {(u-x)_+} - 
\delta(u-x) \ln x \right] = 
\frac{1 - N H(N) + N H\lp N+ \frac{ib} 2 \rp}{N\lp N-\frac{ib} 2\rp}
- \frac{1}{\lp N - \frac{ib} 2 \rp^2}
+\nonumber \\
&&\quad
+ \frac{1}{\lp N - \frac{ib} 2 \rp^2}
= \frac{1 - N H(N) + N H\lp N+ \frac{ib} 2 \rp}{N\lp N-\frac{ib} 2\rp}
= \frac {1+O(N)}{N\lp N - \frac{ib} 2\rp}.
\eea
The inverse Fourier-Mellin of the leading term in this result leads
to the cross-section
\be\label{duinverse}
\frac{\d\sigma_\text{1-leg}}{\d u} (x,u)=
\mathcal M\mathcal F^{-1}\left[\sigma_0 \asb \frac{1}{N\lp N-\frac{ib} 2\rp}
\right]
= \sigma_0 \asb 
\frac{1}{u},
\ee
to be compared to Eq.~\href{du}. Away from the 
endpoint $x=u$, Eq.~\href{du} admits the small $x$ expansion
\be\label{duexp}
\frac{\d\sigma_\text{1-leg}}{\d u} (x,u)= \sigma_0\asb \frac{1}{u-x} = 
\sigma_0\asb \frac{1}{u}\lp 1 + \frac{x}{u} + \lp \frac{x}{u}\rp^2 + ...\rp,
\ee
which implies that Eq.~\href{du} and Eq.~\href{duinverse} are equal up to 
subleading terms.
Indeed, if we integrate Eq.~\href{duexp} we obtain
\be\label{1leg_exp}
\sigma_\text{1-leg} = \sigma_0\asb \lp \ln\frac 1 x + (-x +1) + \frac 1 2
\lp -x^2 + 1\rp + ... \rp.
\ee

This  argument 
shows that at small $x$ 
the $\delta$ function in Eq.~\href{du} cancels an analogous contribution
coming from the endpoint singularity of the plus prescription, thereby
demonstrating that the first term in Eq.~(\ref{du}) is not subleading
in comparison to the second, and that
small $x$ leading terms in rapidity distributions may be unaccompanied
by explicit  $\ln x$ factors. In fact, there can be nontrivial
cancellations among different terms due to endpoint singularity
regularization. The situation is much clearer in Fourier-Mellin-space, where
we can identify all small $x$ enhanced terms as poles in 
$N=0$ or $N=\pm \frac{ib} 2$. 

The emission from the other leg gives an identical contribution 
with only the replacement $u \to 1/u$, which in Fourier space amounts
to $b \to -b$. The full two-leg NLO result in Fourier-Mellin space is thus
\be\label{NLO_2legs_HEFT}
\frac{\d\sigma_\text{NLO}}{\d u}(N,b) = 
\sigma_0 \asb \lp \frac{1}{N\lp N-\frac{ib} 2\rp} 
+\frac{1}{N\lp N+\frac{ib} 2\rp} \rp,
\ee
which in $x$-$u$ space corresponds to the sum of Eq.~(\ref{du}) and the
contribution from the other leg, obtained by letting in $u\rightarrow
1/u$ in Eq.~(\ref{du}).
It is easy to check by inspection
that this NLO result coincides with  the small $x$ limit of the 
full result in~\cite{AnastasiouHiggsNLO}.

We now turn to the NNLO case, for which no closed-form analytic
expression exists. We must  compute the three diagrams of Fig.~\ref{NNLO_pic},
and recall that in the small $x$ limit there are no interferences between
them (see Sect.~\ref{xsres_hadro}).
We start by considering the case of two emissions from the same leg, 
say from the upper leg (see Fig.~\ref{NNLO_pic}-c). 
The result for the differential cross-section is \begin{footnote}{From now on we will neglect all angular 
correlations, which can be shown to be subleading at small $x$. 
Consequently, as in the previous sections we will omit all angular terms like 
$(4\pi)^\epsilon/\Gamma(1-\epsilon)$, as they are always subtracted in the 
$\msb$ scheme. See Appendix~\ref{app_CFP} for details.}\end{footnote}
\be\label{1leg_2g_dsigma}
\d \bar \sigma_{\text{1-leg}}^{\text{2-gluons}}
=\left[
\sigma_0 \frac{\zbar_1 \zbar_2}{x} \delta \lp 1 - 
\frac{\zbar_1 \zbar_2}{x} + \bar \xi_1\rp
\right]
\times
\left[
\asb \frac{d\zbar_2}{\zbar_2} \frac{d\bar \xi_1}{\bar \xi_1^{1+\epsilon}}
\right]
\times
\left[
\asb \frac{d\zbar_1}{\zbar_1} \frac{d\bar\xi_2}{(\bar\xi_1+\bar \xi_2)^{1+\epsilon}}
\right].
\ee
The leading small $x$ behavior in Eq.~\href{1leg_2g_dsigma} comes from the
ordered region $\bar \xi_2\ll \bar \xi_1$, ~$\bar z_i\ll 1$. 
With this in mind we can perform the
$\bar \xi_i$ integrations and the $\msb$ subtraction to obtain
\be
\d \sigma_{\text{1-leg}}^{\text{2-gluons}}
=
\sigma_0 \asb^2 \lp
\frac 1 2 \ln^2\frac 1 x \delta(\zbar_1 \zbar_2 - x)
+
\ln \frac 1 x \frac{1}{(\zbar_1 \zbar_2-x)_+} + 
\left[\frac{\ln(\zbar_1 \zbar_2-x)}{\zbar_1 \zbar_2-x}\right]_+
\rp d\zbar_1 d\zbar_2,
\ee
which leads to the rapidity distribution 
\bea\label{dsigma2_u}
&&\frac{\d \sigma_{\text{1-leg}}^{\text{2-gluons}}}
{\d u} (u,x)= \int \d \sigma_{\text{1-leg}}^{\text{2-gluons}} 
\delta(\zbar_1 \zbar_2 - u)=\nonumber\\
&&\quad=
\sigma_0 \asb^2 \lp
\int_u^1 \frac{d\zbar_1}{\zbar_1}
\rp \lp
\frac 1 2 \ln^2\frac 1 x \delta( u - x)
+
\ln \frac 1 x \frac{1}{(u-x)_+} + 
\left[\frac{\ln(u-x)}{u-x}\right]_+\rp=\nonumber\\
&&\quad=
\sigma_0 \asb^2 
\ln \frac 1 u \lp
\frac 1 2 \ln^2\frac 1 x \delta( u - x)
+
\ln \frac 1 x \frac{1}{(u-x)_+} + 
\left[\frac{\ln(u-x)}{u-x}\right]_+\rp.
\eea

For the sake of  comparison to the resummed result, it is convenient
to consider the Fourier-Mellin transform of Eq.~(\ref{dsigma2_u}). 
Starting from 
Eq.~\href{1leg_2g_dsigma} and performing the $\msb$ subtraction in 
Fourier-Mellin space we obtain
\bea\label{dsigma2_FM}
\frac{\d \sigma_{\text{1-leg}}^{\text{2-gluons}}}
{\d y} (N,b) &=&  \sigma_0 \asb^2 \frac{1}{(N-i b/2)^2} 
\times\nonumber\\
&&\times\left[
\frac{1}{2} \psi^2 (N)+\gamma_E  \psi (N)+\frac{1}{2} \psi
    ^{(1)}(N)+\frac{\pi ^2}{12}+\frac{\gamma_E ^2}{2}\right]=\nonumber \\
&=&\sigma_0 \asb^2 \frac{1}{(N-i b/2)^2}
\frac{1}{N^2}\lp 1 + O(N) \rp.
\eea

Once again, despite naive appearance, 
the three terms in Eq.~(\ref{dsigma2_u}) are all of the same order.
The
small $x$ expansion of Eq.~\href{dsigma2_u} is
\be\label{dsigma2_usmallx}
\frac{\d \sigma_{\text{1-leg}}^{\text{2-gluons}}}
{\d u} (x,u)= \sigma_0 \asb^2 \frac{1}{u}\ln\frac 1 u
\lp \ln \frac{1}{x} + \ln u \rp \lp 1 + O \lp \frac x u \rp \rp,
\ee
or
\be\label{dsigma2_ysmallx}
\frac{\d \sigma_{\text{1-leg}}^{\text{2-gluons}}}
{\d y} (x,y)= 8\sigma_0 \asb^2 y \lp \frac 1 2 \ln \frac 1 x - y \rp
\lp 1 + O\lp x e^{2y}\rp \rp,
\ee
whose Fourier-Mellin transform is  the leading-$N$ term given in the
last step of  
Eq.~\href{dsigma2_FM}.

Having obtained the rapidity distribution for two emissions from the
upper leg (Fig.~\ref{NNLO_pic}-c) it is immediate to obtain the 
result for two emissions from the lower one (Fig.~\ref{NNLO_pic}-a),
through the substitution $y\rightarrow -y$ (i.e.
$u\rightarrow 1/u$ or $b\rightarrow -b$). To complete
the small $x$ NNLO distribution we only need the distribution
with one emission from each leg. In this case the leading small $x$ term
in Fourier-Mellin space reads
\be
\frac{\d \sigma_{\text{2-legs}}^{\text{2-gluons}}}
{\d y} (N,b) 
=2\sigma_0 \asb^2 \frac{1}{N^2}\frac{1}{(N-i b/2)} \frac{1}{(N+i b/2)}.
\ee
Adding everything up we get
\be
\frac{\d \sigma_{\text{NNLO}}}
{\d y} (N,b) = \sigma_0 \asb^2\frac{1}{N^2}
\lp
\frac{1}{N-i b/2}+\frac{1}{N+i b/2}
\rp^2.
\ee

Combining the results of this section, we can write the small $x$ 
rapidity distribution up to NNLO as:
\bea\label{dsigma_HEFT_uptoNNLO}
&&\frac{\d \sigma}{\d y}(N,b) = 
\sigma_0 \left[
1+ \frac{\asb}{N} \lp \frac{1}{N-i b/2}+\frac{1}{N+i b/2}\rp
+\right.\nonumber\\
&&\quad\left.
+ \lp \frac{\asb}{N}\rp^2 \lp \frac{1}{N-i b/2}+\frac{1}{N+i b/2}\rp^2
+O(\as^3)\right].
\eea
Note that letting $b=0$ in this formula we reproduce the small $x$ 
behavior of the inclusive cross-section.

\subsection{Rapidity distribution with finite $m_{\rm top}$}
\label{finitemtop}

As already mentioned, the effective Lagrangian
Eq.~\href{HEFT}
is inadequate at small $x$ and leads to spurious singularities. We now
 turn to the determination of the correct small $x$ rapidity distribution.

As before we start by considering the NLO distribution, i.e. we consider
single gluon emission. For definiteness, we consider first an emission from 
the upper leg.  We obtain:
\be\label{dsigma_1_mt}
\d \bar\sigma_{\text{1-leg}} = \left[
\lp \frac{\pi^3 \as^2 G_F \sqrt{2}}{4}\tau^2
\left|A_1(0,\bar\xi,\tau)\right|^2 \rp \frac{\zbar}{x}
\delta \lp 1 - \frac \zbar x + \bar\xi \rp
\right]\times
\left[ \asb \frac{d\zbar}{\zbar} \frac{d\bar\xi}{\bar\xi^{1+\epsilon}} 
\right],
\ee
where 
\be\label{def_tau}
\tau \equiv 4 m_{\rm top}^2/m_h^2
\ee
 and $A_1$ is the form factor defined
in~\cite{SimoneHiggs}
\begin{footnote}{Note that in
Ref.~\cite{SimoneHiggs} the definition of $\tau$ is different.}\end{footnote}. Again, the hard
part, i.e. the term in the first
square bracket, reproduces the LO result when $\xi\rightarrow 0$, 
$\zbar\rightarrow 1$, as expected from Eq.~\href{onshell}:
\be
\sigma_\text{LO}(\tau) =  \sigma_0 (12 \pi^2 \tau)^2 \left|A_1(0,0,\tau)\right|^2
\equiv \sigma_0(\tau).
\ee
Also, because
\be
\lim_{\tau\rightarrow \infty} \tau A_1(\xi,0,\tau) = 
\frac {1}{12 \pi^2}
\ee
in the heavy top approximation Eq.~\href{dsigma_1_mt} reproduces our
previous result Eq.~\href{higgs_NLO}. 

The form factor $A_1$ acts as 
a cut-off for the $\bar\xi$ integration at large $\bar\xi$, thus removing the
spurious small $x$ singularities of the effective theory. To see how
this works, we consider the rapidity distribution in Fourier-Mellin space.
Starting from Eq.~\href{dsigma_1_mt} we obtain:
\be
\frac{\d\bar \sigma_\text{1-leg}}{\d y} (N,b)= 
\sigma_0 (12 \pi^2 \tau)^2
\lp \frac{\asb}{N-i b/2}\rp 
\int_0^\infty \frac{d\bar\xi}{\bar\xi^{1+\epsilon}}
\frac{\left|
A_1(0,\bar\xi,\tau)
\right|^2}
{\lp 1 + \bar\xi \rp^N},
\ee
where we have used the delta function to perform the $N-$Mellin 
transform. Without the form factor $A_1$ (i.e. in the effective theory)
the $\bar\xi$ integral would be UV divergent at $N=0$. This leads to a
spurious $1/N$ behavior of the integral. In the full theory this problem
is no longer there: the integral is cut off by $A_1$, hence at small $x$
we can safely let $N=0$:
\be
\frac{\d \bar\sigma_\text{1-leg}}{\d y} (N,b)= 
\sigma_0 (12 \pi^2 \tau)^2
\lp \frac{\asb}{N-i b/2}\rp 
\int_0^\infty \frac{d\bar\xi}{\bar\xi^{1+\epsilon}}
\left|
A_1(0,\bar\xi,\tau)
\right|^2.
\ee

The $\bar\xi$ integral now is well-behaved in the ultraviolet, 
but it still contains 
a collinear singularity. To extract it, we simply integrate by parts:
\bea
&&\int_0^\infty 
\frac{d\bar\xi}{\bar\xi^{1+\epsilon}}
\left|
A_1(0,\bar\xi,\tau)
\right|^2 =\nonumber\\
&&\quad=
 -\frac 1 \epsilon
\left\{
\left[
\left|A_1(0,\bar\xi,\tau)\right|^2 \bar\xi^{-\epsilon}\right]_{\bar\xi=0}^{\bar\xi\rightarrow\infty}
- \int_0^\infty d\bar\xi \bar\xi^{-\epsilon} \frac{d \left|A_1(0,\bar\xi,\tau)\right|^2}
{d\bar\xi}
\right\}\nonumber\\
&&\quad= -\frac{1}{\epsilon}
\left\{
\left|A_1(0,0,\tau)\right|^2 + \epsilon \int_0^\infty d\bar\xi \ln \bar\xi 
\frac{d \left|A_1(0,\bar\xi,\tau)\right|^2}
{d\bar\xi}+ O(\epsilon^2) 
\right\}, \nonumber \\
\eea
where we have used
\be
\lim_{\bar\xi \rightarrow 0, \infty} \left|A_1(0,\bar\xi,\tau)\right|^2 \bar\xi^{-\epsilon}=0
\label{fflimit}
\ee
since $\epsilon < 0$. We can thus write the $\msb$ subtracted rapidity
distribution as
\be\label{rap_1leg_mt}
\frac{\d \sigma_{\text{1-leg}}}{\d y}(N,b) = 
\sigma_0(\tau) \frac{\asb}{N-i b/2} c_1(\tau),
\ee
where we have defined
\be\label{c1}
c_1(\tau) \equiv - \frac{1}{\left|A_1(0,0,\tau)\right|^2}
\int_0^\infty d\bar\xi \ln \bar\xi 
\frac{d \left|A_1(0,\bar\xi,\tau)\right|^2}
{d\bar\xi}.
\ee

Numerical results for $c_1(\tau)$ are tabulated in Tab.~\ref{tablecoeff}.
Note that the result Eq.~\href{rap_1leg_mt} does not 
depend on $N$ and $b$ separately, but only
on the combination $N-i b/2$. This implies that in $(x,y)$ space the result
has a rapidity delta function:
\be
\frac{\d \sigma_{\text{1-leg}}}{\d y}(x, y) = 
\sigma_0(\tau) c_1(\tau) \asb \delta \lp y - \frac 1 2 \ln \frac 1 x \rp.
\label{1legxs}
\ee

To obtain the result for emission from the lower leg it is sufficient to
symmetrize this result. The full small $x$ NLO result thus is
\be\label{rap_NLO_mt}
\frac{\d \sigma_{\text{NLO}}}{\d y} (x,y) = 
\sigma_0(\tau) c_1(\tau) \asb \left[  \delta \lp y - \frac 1 2 \ln \frac 1 x \rp +
\delta \lp y + \frac 1 2 \ln \frac 1 x \rp \right],
\ee
or in Fourier-Mellin space
\be\label{rap_NLO_mtfm}
\frac{\d \sigma_{\text{NLO}}}{\d y} (N,b) = 
\sigma_0(\tau) c_1(\tau) \asb 
\lp \frac{1}{N-i b/2} + \frac{1}{N+i b/2} \rp.
\ee

Note that the small $x$ behavior of the  rapidity distribution with finite
top mass is 
completely different from that  in the effective theory: in the
former case the rapidity is forced to sit at its endpoints, while in the latter
we have a flat rapidity distributions (see Eq.~\href{duinverse} and
recall $dy =  du/(2u)$). Note also that the integration of
Eq.~\href{rap_NLO_mt} leads to a constant, in contrast with
the spurious $\ln x$ behavior of the effective theory.

We can compare also the NNLO distributions. As in the previous subsection, 
we start from two emissions from the upper leg. In this case 
the small $x$ rapidity distribution in Fourier-Mellin space is
\be\label{rap_2g_mt}
\frac{\d\bar \sigma_\text{1-leg}^\text{2-gluons}}{\d y}(N,b) = 
\sigma_0 (12\pi^2\tau)^2
\lp \frac {\asb}{N-i b/2} \rp^2
\int_0^\infty \frac{d\bar\xi}{\bar\xi^{1+\epsilon}}
\frac{
\left|A_1(0,\bar\xi,\tau)\right|^2
}
{(1+\bar\xi)^N}
\lp -\frac 1 \epsilon \frac{1}{\bar\xi^{\epsilon}} + \frac{1}{\epsilon}\rp,
\ee
where we have regulated just one of the two collinear singularities coming
from the extra gluon emission. To deal with Eq.~\href{rap_2g_mt} we first note
that we can safely let $N=0$ in the $\bar\xi$ integral as before. To deal with
the remaining collinear singularity, 
we just integrate by parts as in the NLO case. The $\msb$ result is then
\be
\frac{\d \sigma_\text{1-leg}^\text{2-gluons}}{\d y}(N,b) = 
\sigma_0(\tau) c_2(\tau) \lp \frac{\asb}{N-i b/2}\rp^2,
\label{2emres}
\ee
with
\be\label{c2}
c_2(\tau) \equiv - \frac{1}{\left|A_1(0,0,\tau)\right|^2}
\int_0^\infty d\bar\xi
\frac{\ln^2 \bar\xi}{2} \frac{d \left|A_1(0,\bar\xi,\tau)\right|^2}{d \bar\xi}.
\ee
Numerical results for $c_2$ are tabulated in Tab.~\ref{tablecoeff}.

\begin{table}
\centering
\begin{tabular}{|c||c||c|c|}
\hline
$\tau$ & $c_1$ & $c_2$ & $c_{1,1}$ \\\hline
\hline
 1.5000 &   0.4869 &   1.6119  &  0.7752 \\\hline
    2.0000 &    0.8372 &    1.8540 &    1.2129 \\\hline
    2.5000 &  1.0923 &    2.1056 &    1.6925 \\\hline
    3.0000 &    1.2942 &    2.3500 &    2.1670 \\\hline
    3.5000 &    1.4616 &    2.5831 &    2.6237 \\\hline
    4.0000 &    1.6046 &    2.8042 &    3.0596 \\\hline
    4.5000 &    1.7297 &    3.0141 &    3.4746 \\\hline
    5.0000 &    1.8407 &    3.2134 &    3.8699 \\\hline
    5.5000 &    1.9406 &    3.4031 &    4.2470 \\\hline
    6.0000 &    2.0314 &    3.5841 &    4.6073 \\\hline
    6.5000 &    2.1146 &    3.7571 &    4.9522 \\\hline
    7.0000 &    2.1914 &    3.9230 &    5.2831 \\\hline
    7.5000 &    2.2627 &    4.0822 &    5.6011 \\\hline
    8.0000 &    2.3292 &    4.2353 &    5.9073 \\\hline
    8.5000 &    2.3916 &    4.3829 &    6.2026 \\\hline
    9.0000 &    2.4502 &    4.5253 &    6.4877 \\\hline
    9.5000 &    2.5058 &    4.6630 &    6.7635 \\\hline
   10.0000 &    2.5582 &    4.7963 &    7.0306 \\\hline
   10.5000 &    2.6081 &    4.9255 &    7.2896 \\\hline
   11.0000 &    2.6556 &    5.0509 &    7.5411 \\\hline
   11.5000 &    2.7010 &    5.1724 &    7.7854 \\\hline
   12.0000 &    2.7444 &    5.2908 &    8.0231 \\\hline
\end{tabular}
\caption{Numerical values for the NLO coefficient $c_1$ Eq.~\href{c1}
and for the NNLO coefficients $c_2$ Eq.~\href{c2} and
$c_{1,1}$ Eq.~\href{c11}.}
\label{tablecoeff}
\end{table}

The results with two emissions from the lower leg is trivially obtained
by performing the replacement $b\rightarrow -b$ in Eq.~(\ref{2emres}).
In the 
nontrivial case of one emission from each leg the differential
cross-section is
\bea
&&\d \bar\sigma_\text{2-legs}^\text{2-gluons} =
\left[
\sigma_0 (12 \pi^2 \tau)^2 
2\left|A_1(\xi,\bar \xi,\tau) \cos\theta + A_3(\xi,\bar \xi,\tau)
\sqrt{\xi\bar \xi}
\right|^2 
\frac{\zbar z}{x} 
\right.
\times\nonumber\\
&&\quad\times
\left. 
\delta\lp 1- \frac{\zbar z}{x} + \xi + \bar \xi + 2 \sqrt{\xi \bar \xi}
\cos\theta \rp \frac{d\theta}{2\pi} \right]
\left[\asb \frac{d\zbar}{\zbar} \frac{d \bar\xi}{\bar\xi^{1+\epsilon}}
\right]
\left[\asb \frac{dz}{z} \frac{d \xi}{\xi^{1+\epsilon}}
\right],
\label{twolegsep}
\eea
where $A_3$ was defined in Ref.~\cite{SimoneHiggs}. 

Note that if we let $\xi=0$ in the first square bracket of
Eq.~(\ref{twolegsep})  and
perform the angular integration we recover the one-leg result 
Eq.~\href{dsigma_1_mt}. 
From the differential cross-section Eq.~(\ref{twolegsep}) we can easily
write down the rapidity distribution in Fourier-Mellin space
\bea
\frac{\d\bar \sigma_\text{2-legs}^\text{2-gluons}}{\d y}(N,b)&=&
\sigma_0 (12\pi^2 \tau)^2 
\lp\frac{\asb}{N-i b/2}\rp
\lp\frac{\asb}{N+i b/2}\rp\nonumber\times\\
&\times&
\int \frac{d \xi}{\xi^{1+\epsilon}}
\frac{d \bar \xi}{\bar \xi^{1+\epsilon}}
\frac{\left|A_1(\xi,\bar \xi,\tau) \cos\theta + A_3(\xi,\bar \xi,\tau)
\sqrt{\xi\bar \xi}\right|^2}{(1+\xi+\bar \xi+2 \sqrt{\xi \bar \xi} 
\cos\theta)^N} \frac{d\theta}{\pi}.
\eea

As in the previous case we can let $N=0$ in the integral and
perform the $\msb$ subtraction by integrating by parts. The final 
subtracted result now reads
\be
\frac{\d \sigma_\text{2-legs}^\text{2-gluons}}{\d y}(N,b) = 
\sigma_0(\tau) c_{1,1}(\tau)
\lp\frac{\asb}{N-i b/2}\rp
\lp\frac{\asb}{N+i b/2}\rp
\ee
with
\be\label{c11}
c_{1,1}(\tau) \equiv
\frac{1}{\left|A_1(0,0,\tau)\right|^2}
\int_0^\infty d\xi
\int_0^\infty d\bar \xi \left[
\ln \xi \ln \bar \xi \frac{\partial^2 \left| A_1(\xi,\bar \xi,\tau)\right|^2}
{\partial\xi \partial\bar \xi}
+2 \left|A_3(\xi,\bar \xi,\tau)\right|^2
\right].
\ee
Numerical values for $c_{1,1}$ are tabulated in Tab.~\ref{tablecoeff}.

Adding up all contribution up to NNLO  we obtain the following small $x$ 
rapidity distribution:
\bea
\frac{\d\sigma}{\d y}(N,b) &=&
\sigma_0(\tau)\Bigg\{1 + c_1(\tau)\asb 
\left[
\frac{1}{N-i b/2} + \frac{1}{N+i b/2}
\right] \nonumber\\
&+&  \asb^2 \Bigg[
 c_2(\tau)
\lp\lp \frac{\asb}{N-i b/2}\rp^2+ \lp \frac{\asb}{N+i b/2}\rp^2
\rp  \nonumber \\ && + 
 c_{1,1}(\tau)\lp \frac{\asb}{N-i b /2}\rp
\lp \frac{\asb}{N+i b /2}\rp \Bigg]
+ O(\as^3)
\Bigg\}.
\eea

\section{Resummation of the Higgs rapidity distribution}
\label{higgsres}

In this section we use Eq.~\href{rap_res_msb}
(or, equivalently, Eq.~(\ref{catanilikeresum})) to
perform the small $x$ resummation of the Higgs rapidity distribution,
both in the effective theory and for finite top mass. We will check
that that up to NNLO the resummed results agree with the explicit
computations presented in the previous section.

\subsection{Pointlike effective interaction}
\label{pointlike}

The resummation of the inclusive cross-section in the limit $m_{\rm
top}\to\infty$ was first performed
in Ref.~\cite{HautmannHiggs}. The  resummed result for the gluon-gluon sub-process  in
Mellin space is
\bea
\sigma_{gg}(N)
&=&\frac{\sigma_0}{1-\frac{2\gamma}{N}}R_{\msb}^2(\gamma)=
R_{\msb}^2(\gamma) \sigma_0 \lp 1 + \frac{2\gamma}{N} + 
\lp \frac{2\gamma}{N}\rp^2+...\rp
\label{haut_incl}\eea
where $\gamma=\gamma\left(\frac{\alpha_s}{N}\right)$ is the LL$x$
anomalous dimension.
Because of the  simple kinematic relation~\href{y_twolegs}
between $z$ and $y$, the only further piece of information needed for
the resummation is the off-shell cross-section with the contributions
from emissions coming from the two legs kept separate. 
This is given by~\cite{HautmannHiggs}
\be\label{haut_m_mbar}
\sigma_{gg}(N)=\frac{\sigma_0}{1-M-\bar M}
R_{\msb}\left(M\right) R_{\msb}\left(\bar M\right)\Bigg|_{M=\bar M=\gamma\left(\frac{\alpha_s}{N}\right)},
\ee
where the two variables $M$, $\bar M$ correspond to radiation from
either leg. Using  Eq.~\href{catanilikeresum}, we 
immediately get
\be\label{higgs_res_pl}
\frac{\d\sigma_{gg}}{\d y}(N,b) =\frac{\sigma_0 R_{\msb}
\lp\gamma\lp\frac{\as}{N -i b/2} \rp \rp R_{\msb}
\lp\gamma\lp\frac{\as}{N +i b/2} \rp \rp}
{1-\frac{1}{N}\gamma \lp \frac{\as}{N-ib/2}\rp-
\frac{1}{N}\gamma \lp \frac{\as}{N+ib/2}\rp}.
\ee
Note that, in accordance with Eq.~\href{catanilikeresum}, the argument
$N$ in the anomalous dimension is  shifted by $\pm ib$, but  the
intrinsic 
$N$ dependence of the coefficient
function is not.

Recalling that $R_{\msb} = O(\as^3)$ it is immediate to check that the 
resummed result  agrees with the fixed-order results up to NNLO 
obtained in Sect.~\ref{mtopinfty} Eq.~\href{dsigma_HEFT_uptoNNLO}.~\footnote{It follows that these
  results cannot be used to check a recent
  claim~\cite{Kirschner:2009qu} that the standard result of $R$
  from Refs.~\cite{CataniHQ,Catani:1993ww} is incorrect.}
Expanding Eq.~\href{higgs_res_pl} up to NNLO and performing
the inverse Fourier-Mellin transform, the result in terms of 
the partonic rapidity $y$ is
\bea \label{higgs_res_xsp_exp}
\frac{\d\sigma_{gg}}{\d y}(x,y) &=& 
\sigma_0 \Bigg \{
\frac{\delta(1-x)}{2} \left[ \delta\lp y - \frac 1 2 \ln \frac 1 x \rp+
\delta \lp y - \frac 1 2 \ln x \rp \right]  \nonumber\\ &&+ 2 \asb
+ 4 \asb^{2}
\lp \frac {\ln^2 x}{4}-y^2 \rp + O(\as^3) \Bigg \}.
\eea

\subsection{Finite top mass}
\label{mtfin}

We now consider the case of  finite top mass. The resummation
of the inclusive cross-section has been performed in
Ref.~\cite{SimoneHiggs}. The result was written in terms of a double
Mellin transform
\bea\label{higgs_resolv_res}
\sigma_{gg}(N) &=&  \sigma_0 (12\pi^2)^2 \tau^2
R_{\msb}(M)R_{\msb}(\bar M)
M \bar M \times\nonumber \\ &&
\times\int_0^\infty d\xi \xi^{M-1}\int_0^\infty d\bar \xi \bar \xi^{\bar M-1}
\left[ \left| A_1\right|^2 + 2 \xi \bar \xi \left|A_3\right|^2
\right]\Bigg|_{M=\bar M=\gamma\left(\frac{\alpha_s}{N}\right)},
\eea
where as above  $M$ and $\bar M$ correspond to radiation from either
of the two legs.
Expanding Eq.~\href{higgs_resolv_res} in powers of $M$ and $\bar M$
one obtains the coefficient of the LL$x$ singularity to any desired
order in terms of double integrals over $\xi$ and $\bar \xi$, which
have to be evaluated numerically.
Note that now in Eq.~\href{higgs_resolv_res} all the $N$ dependence comes
from $\gamma(\as/N)$, since in the small $x$ limit the
form factors $A_i$ are $N-$independent as one expects for a non-pointlike interaction.

 It is immediate to obtain
the rapidity distribution:
\bea\label{higgs_dy_resv}
\frac{\d\sigma_{gg}}{\d y}(N,b) &=& \sigma_0 (12 \pi^2)^2 \tau^2
R_{\msb} ({M}_b)R_{\msb} (\bar{M}_b)
 M_b \bar{M}_b \\ \nonumber &&
\int_0^\infty d\xi \xi^{M_b-1}\int_0^\infty d\bar \xi
\bar \xi^{\bar{M}_b-1}
\left[ \left| A_1\right|^2 + 2 \xi \bar \xi \left|A_3\right|^2
\right],
\eea
where we have defined
\be
M_b \equiv \gamma \lp \frac{\as}{N-\frac{ib} 2} \rp;
\quad
\bar{M}_b \equiv \gamma \lp \frac{\as}{N+\frac{ib} 2} \rp.
\ee
Integrating Eq.~\href{higgs_dy_resv} by parts we can make 
its perturbative expansion explicit: 
\bea\label{rap_mt_res}
&&\frac{\d\sigma_{gg}}{\d y}(N,b) = 
\sigma_0(\tau) R_{\msb} ({M}_b)R_{\msb} (\bar{M}_b)
\times\nonumber\\ 
&&\quad\times\left[\sum_{i\ge 0}\lp M_b^i + \bar{M}_b^i \rp c_i(\tau)
+ \sum_{j,k > 0} M_b^j \bar{M}_b^k c_{j,k}(\tau)
\right]
\eea
with 
\bea
c_i(\tau) &\equiv& -\frac{1}{\left|A_1(0,0,\tau)\right|^2}
\int_0^\infty d\xi \frac{\ln^i \xi}{i!} 
\frac{d\left|A_1(\xi,0,\tau)\right|^2}{d\xi}\nonumber\\
c_{j,k}(\tau) &\equiv&
\frac{1}{\left|A_1(0,0,\tau)\right|^2}
\int_0^\infty d\xi
\int_0^\infty d\bar \xi \left[
\frac{\ln^j \xi}{j!}
\frac{ \ln^k \bar \xi}{k!}
 \frac{\partial^2 \left| A_1(\xi,\bar \xi,\tau)\right|^2}
{\partial\xi \partial\bar \xi}\nonumber \right.\\
&&+2 \left.\frac{\ln^{j-1}\xi}{(j-1)!}\frac{\ln^{k-1}\bar \xi}{(k-1)!}
\left|A_3(\xi,\bar \xi,\tau)\right|^2
\right]. 
\eea
Integrating Eq.~\href{rap_mt_res} 
we obtain the right single logarithmic behavior of the total cross-section. 

For comparison with the explicit results of the previous section, we 
write the $(x,y)$ resummed rapidity distribution up to NNLO:
\bea\label{resNNLOx}
&&\frac{\d\sigma_{gg}}{\d y}(x,y) = \sigma_0(\tau) 
\left\{
\frac{\delta(1-x)}{2} \left[
\delta\lp y-\frac{1}{2}\ln x\rp+\delta\lp y+\frac{1}{2}\ln x\rp\right]
+\right.\nonumber\\
&&\quad
+\asb c_1(\tau)\left[
\delta\lp y-\frac{1}{2}\ln x\rp+ \delta\lp y+\frac{1}{2}\ln x\rp
\right]+\nonumber\\
&&\quad\left.
+\asb^2 \left[
\left[ \delta
\lp y - \frac 1 2 \ln x \rp+\delta
\lp y + \frac 1 2 \ln x \rp
\right]  \ln \frac{1}{x} c_{2}(\tau)
+c_{1,1}(\tau)
\right]
\right\},
\eea
which agrees with our previous findings. 
The pattern of Eq.~\href{resNNLOx} persists at higher orders: there 
are two endpoint contributions coming from emissions from just one leg,
plus a bulk term coming from emissions from both legs.

So far  we have only considered  the gluon-gluon partonic sub-process.
However, beyond LO other channels open up.  As discussed in
Sect.~\ref{xsres_hadro}, 
the high energy behavior of the quark initiated sub-processes can be
straightforwardly computed from the gluon-gluon case thanks to 
color-charge relations. The LL$x$ behavior for the inclusive Higgs
production in the different partonic channels have been explicitly
computed to NNLO in~\cite{SimoneHarlander}. The generalization 
to the rapidity distributions is not difficult. Up to NNLO
we have
\bea
&&\frac{\d\sigma_{qg}}{\d y}(x,y) = \sigma_0(\tau) \frac{C_F}{C_A}
\Bigg \{ \asb \left[ \delta
\lp y - \frac 1 2 \ln x \rp+ \delta
\lp y + \frac 1 2 \ln x \rp
\right]
\frac{c_1(\tau)}{2} +\label{qgNLO} \nonumber\\  
&&\quad
+ \asb^2
\left[ \delta
\lp y - \frac 1 2 \ln x \rp + \delta
\lp y + \frac 1 2 \ln x \rp
\right]  \ln \frac{1}{x} 
\frac{c_{2}(\tau)}{2} +c_{1,1}(\tau) + O(\as^3) \Bigg \};  \nonumber\\  \\
&&\frac{\d\sigma_{q_iq_j(\bar q_j)}}{\d y}(x,y) = \sigma_0(\tau)
\Bigg \{\asb^2 \left(\frac{C_F}{C_A}\right)^2 c_{1,1}(\tau) + O(\as^3)
\Bigg \}.
\label{qqNNLO}\eea

\subsection{Matching to the effective theory}
\label{matching}

As mentioned in the introduction, resummed results are useful not only
for the sake of all-order resummed phenomenology, but also as a way of
obtaining information on higher order perturbative terms. This is
particularly interesting in the case of Higgs production, where
results are most easily obtained in the $m_{\rm top}\to\infty$ limit,
which however fails at small $x$. It is then possible to improve the 
$m_{\rm top}\to\infty$ result by correcting its small $x$ behavior
with its exact finite $m_{\rm top}$ form extracted from
resummation. For the inclusive NNLO cross-section this was done
in Ref.~\cite{SimoneHiggsProc}. A more ambitious task is to construct
an accurate approximation to the full finite $m_{\rm top}$ result by
matching the small $x$ behavior from resummation to an expansion in
powers of $1/m_{\rm top}$, which at the  inclusive NNLO level was 
done in Ref.~\cite{SimoneHarlander}.

Here we perform a  matched
determination akin to that of Ref.~\cite{SimoneHiggsProc} 
for the NLO rapidity distribution, by combining the small $x$
behavior Eqs.~(\ref{rap_NLO_mt}, \ref{qgNLO}) with the analytic expression in the
$m_{\rm top}\to\infty$ limit from Ref.~\cite{AnastasiouHiggsNLO}.
 This quantity has also been obtained by
numerical integration of a fully differential expression in
Ref.~\cite{hpro}, with which we will compare our results.
In Ref.~\cite{AnastasiouHiggsNLO} the rapidity distribution is
parametrized by the variables $z$ and $y$ where $z$ coincides with  the variable
that we have called $x$ throughout this paper (compare in particular
Eq.~(\ref{rap_def}))  while $y$
should not be confused with the rapidity, and coincides with the
variable $w$ defined as
\be\label{def_w}
w \equiv \frac{u-x}{(1-x)(1+u)}
\ee
in terms of $x$ and $u$ Eq.~(\ref{udef}). 

The matched result is constructed as  
follows:
\be
\frac{\d \sigma_{ij}}{\d w}=\left\{\begin{array}{c}
\frac{d \sigma_{ij}}{d w}\big|_{\rm eff}\phantom{+
\frac{d \sigma_{ij}}{d w}\big|_{\rm matching}}\qquad x>x_\text{match}\\
\frac{d \sigma_{ij}}{d w}\big|_{\rm eff}+
\frac{d \sigma_{ij}}{d w}\big|_{\rm matching}\qquad x\le x_\text{match}
\end{array} \right. 
\label{matchedres}
\ee
where $ij$=$gg$ or $qg$, by $\frac{\d \sigma_{ij}}{\d
  w}\big|_{\rm eff}$ we denote the result in the
the heavy top limit
as given in Eqs.~(24)-(25) of Ref.~\cite{AnastasiouHiggsNLO}, and 
$\frac{\d \sigma_{ij}}{\d w}\big|_{\rm matching}$ is a
matching term which subtracts the spurious double log small $x$
behavior from the heavy top result, and it replaces it with the
correct small $x$ behavior for finite $m_{\rm top}$. 

The matching term in turn is determined by noting that at NLO at the
inclusive level the
spurious double logs correspond to terms which in Mellin space behave as
either a simple or a double $N$ pole (i.e. as functions of $x$ which are
either constant or grow as $\ln x$ as $x\to0)$. Thus, the matching terms are
\bea
\lp\frac{\d \sigma_{gg}}{\d w}\rp_{\text{matching}} 
&=&
-\left[
3 \frac{x^2-1}{\left[ w(x-1)+1 \right]
\left[ w(x-1)-x \right]} -3(2-w(1-w))
\right] + \nonumber \\
&+& 
3 c_1(\tau) \left[ \delta(1-w) + \delta(w) \right],\label{matchgg}\\
\lp\frac{\d \sigma_{qg}}{\d w}\rp_{\text{matching}} 
&=&
-\left[
2 \frac{x^2-1}{\left[ w(x-1)+1 \right]
\left[ w(x-1)-x \right]} -1
\right] + \nonumber \\
&+& 
2 c_1(\tau) \left[ \delta(1-w) + \delta(w) \right].\label{matchqg}
\eea
where the first line of Eqs.~(\ref{matchgg},\ref{matchqg})
subtracts all contributions to the heavy top results
(respectively Eq.~(25) and Eq.~(24) of Ref.~\cite{AnastasiouHiggsNLO})
which upon integration lead to contributions to the inclusive result
which either grow or go to a constant at small $x$. The second line of
both expressions adds back the correct single-log behavior as given
respectively in Eq.~(\ref{rap_NLO_mt}), and Eq.~(\ref{qgNLO}).
Note that the first term on the right-hand side of
Eqs.~(\ref{matchgg}, \ref{matchqg}) is the leading double-log
contribution computed in Sect.~\ref{pointlike}, Eq.~(\ref{duexp}),
symmetrized, expressed in terms of $w$ and multiplied by an
appropriate Jacobian; when expressed in terms of the rapidity $y$ this
contribution is a flat rapidity distribution, as already noticed in
Eq.~(\ref{rap_NLO_mtfm}).  The second
subtraction term on the right-hand side of
Eqs.~(\ref{matchgg}, \ref{matchqg}) is NLL$x$ hence it is not
determined by LL$x$ resummation 
and we have extracted it from Ref.~\cite{AnastasiouHiggsNLO} directly.
We will choose for $x_\text{match}$  the same matching value as used
in Refs.~\cite{SimoneHiggs,SimoneHiggsProc}. This ensures that our result
reproduces the inclusive one once it has been integrated over the 
rapidity range.

\begin{figure}
\begin{center}
\epsfig{file=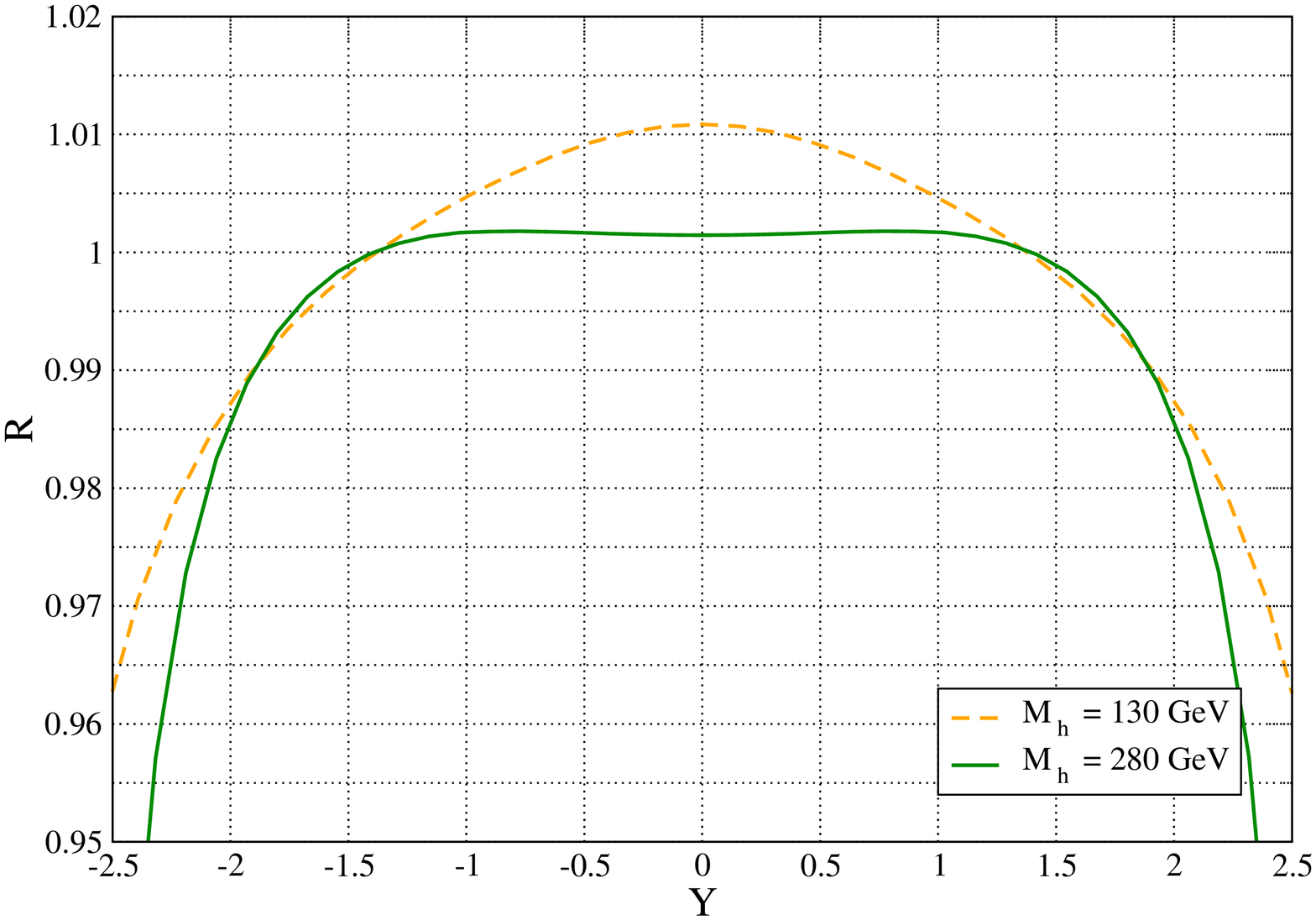, width=0.35\textwidth, angle=-90}
\epsfig{file=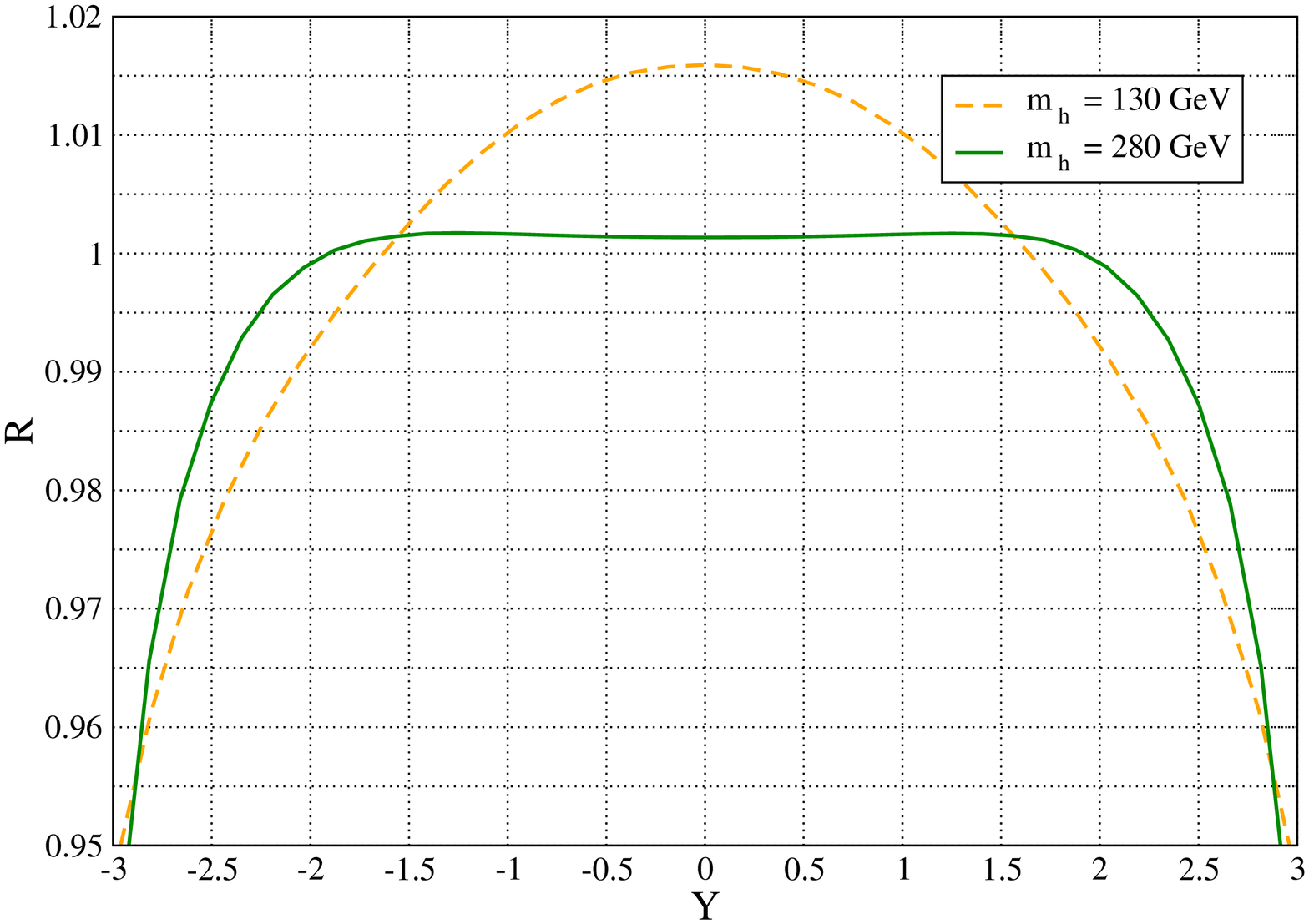, width=0.35\textwidth, angle=-90}
\end{center}
\caption{The ratio of the rapidity distribution for 
Higgs production computed up to NLO  using the matched expressions
Eqs.~(\ref{matchedres}-\ref{matchqg})  to the result in the heavy top
limit of Ref.~\cite{AnastasiouHiggsNLO}. The ratio is rescaled by the
corresponding ratio of total cross-sections.
Results are shown for
the LHC at $7$~TeV (left) and $14$~TeV (right); the dashed yellow 
curves are for $m_h=130$~GeV and the solid green ones for $m_h=280$~GeV.}
\label{plot:ratio1}
\end{figure}

We have computed the full hadronic  rapidity distribution up to NLO,
using for the NLO term 
the matched results Eqs.~(\ref{matchedres}-\ref{matchqg}). The
hadron-level result is given by 
\be
\frac{\d \sigma}{\d Y} = \int_0^1 dw \int_{x_h}^1 dx \frac{\d \sigma}{\d w}
\mathcal L(x,w,Y),
\label{hadrapw}
\ee
where the partonic luminosity $\mathcal L$ is defined as
\be
\mathcal L(x,w,Y) \equiv \frac{x_h}{x^2} 
f\lp \frac{\sqrt{x_h}e^Y}{\sqrt{x/u}}\rp
f\lp \frac{\sqrt{x_h}e^{-Y}}{\sqrt{x/u}}\rp,
\label{lumiw}
\ee
and $u$ is obtained by inverting Eq.~\href{def_w}. 
Results for two different values of the Higgs mass
($m_h=130$~GeV and $m_h= 280$~GeV), at the LHC with $\sqrt{S}=7$~TeV 
and $\sqrt{S}=14$~TeV are shown
in Fig.~\ref{plot:ratio1}; we have used NNPDF2.0 
parton distributions~\cite{nnpdf2.0} (PDF uncertainties are
not shown). We plot the ratio of the NLO matched result to the large top
mass result of Ref.~\cite{AnastasiouHiggsNLO}, divided by the
corresponding ratio of total cross-sections, so that only shape
differences are shown in the plot.
 These plots confirm
the conclusion of Ref.~\cite{hpro} that corrections to the NLO
rapidity
distribution due to finite-mass effects are below 5\%. We also
determine the precise  shape of the correction, which 
could not be determined in Ref~\cite{hpro} because of insufficient
numerical accuracy. In comparison to that reference, we also seem to
find a somewhat more noticeable correction in the central rapidity region,
though of course the effect on the total cross-section remains quite
small. For heavier Higgs masses the correction to the shape becomes smaller.

More refined matching procedures might be constructed, for example by
taking the value of $x_\text{match}$ to depend on $y$. However,
the subtraction term (first line on the right-hand side of
Eqs.~(\ref{matchgg}, \ref{matchqg}) turns out to have a very small
impact on the shape correction, at least for the values of the Higgs mass
and energy considered here. 
This can be understood in part as a consequence of the fact that the
leading double logarithmic term (first term on the right-hand side of
Eqs.~(\ref{matchgg}, \ref{matchqg}) is flat in rapidity, as mentioned
above, so it does not affect the shape.

A similar matched procedure  cannot be carried out to NNLO because
an analytic  expression of the Higgs rapidity distribution to 
NNLO does not exist even in the heavy top limit, to the best of our knowledge.
It is possible nevertheless to provide a rough  estimate (more likely
an upper bound) of the impact of finite top mass corrections on the
NNLO cross-section by computing the $K$--factor
\be
K = \left (\frac{\d \sigma^{\rm LO}}{\d Y}+\frac{\d \sigma^{\rm NLO}}
{\d Y}+\frac{\d \sigma^{\rm NNLO}_{{\rm LL}x}}{\d Y}\right)_{\rm matched}
\Bigg/
\left (\frac{\d \sigma^{\rm LO}}{\d Y}+\frac{\d \sigma^{\rm NLO}}
{\d Y}+\frac{\d \sigma^{\rm NNLO}_{{\rm LL}x}}{\d Y}\right)_{\rm{eff}}
\ee  
where at NNLO level only the contribution from the small $x$ tail is
included, in turn approximated in each case by its LL$x$ behavior, namely
\be
\frac{\d \sigma^{\rm NNLO}_{{\rm LL}x}}{\d Y}
\equiv \int_0^1 dw \int_{x_h}^{x_\text{match}} dx 
\frac{\d \sigma^{\rm NNLO}_{{\rm LL}x}}{\d w}
\mathcal L(x,w,Y)
\ee
with  $\d \sigma^{\rm NNLO}_{{\rm LL}x}/\d w$ given by 
the NNLO term in Eq.~\href{resNNLOx} for the matched case and by
the NNLO term of Eq.~\href{higgs_res_xsp_exp} 
for the effective (heavy top)  case, in  each case  multiplied by the
appropriate Jacobian. 
We found that at the LHC at 7~TeV the finite mass corrections are well 
below 1\%, while at 14~TeV they are larger but still below 2\% in the 
full rapidity range. This suggests that the use of the effective 
theory is fully justified for rapidity distributions also beyond NLO.

\section{Conclusions and Outlook}
\label{conclusions}

In this paper we have accomplished for the first time small $x$
resummation
of a differential cross-section (as opposed to a fully inclusive
one). 
This has been done by
extending to differential rapidity distributions the so-called high
energy or $k_T$ factorization, which expresses LL$x$ cross-sections in
terms of standard partonic cross-sections, but computed with incoming
off-shell gluons. The result  has been made possible by exploiting the duality
which relates LL$x$ and LL$Q^2$ evolution equations, to re-express
high energy factorization in terms of standard collinear
factorization, which in turns is straightforwardly extended to
rapidity distributions.

Our final resummation prescription is quite close to the
well-established one derived long ago for inclusive
cross-sections~\cite{CataniHQ}. Indeed, it is based on evaluating
the LO
rapidity distribution for the relevant process but with incoming
off-shell gluons. As in the inclusive case, the
leading high energy behavior is found by means of a pole
approximation in Mellin space, whereby the Mellin variable which is
conjugate to the gluon virtuality is identified with the LL$x$
anomalous dimension. However, one must now also
take a Fourier transform with respect to rapidity, and the argument of
LL$x$ anomalous dimensions undergoes an impact-parameter dependent
shift in the complex plane of the variable which is Mellin conjugate
to the scaling variable $x$. 

As first application, we have performed a resummation of the rapidity
distribution for Higgs production in gluon-gluon fusion. We have computed the
small $x$ limit of the rapidity distribution to NNLO, both in the
heavy top mass approximation and for finite $m_{\rm top}$ (only the
$m_{\rm top}\to\infty $ NLO result being already available in closed
form). 
We have shown the way these
expressions are related to the results obtained by expanding to finite
order the resummation formula and inverting the Fourier-Mellin transform, and we
have seen that this  involves nontrivial cancellations between terms
which at the resummed level have different origins, thereby providing
a nontrivial consistency check on the resummation procedure. As a first
example of application we have constructed an approximate analytic
expression for the NLO rapidity distribution with finite top mass,
which appears to be in agreement with the numerical results of
Ref.~\cite{hpro}, but provides a prediction which is not affected
by problems of numerical accuracy.

The potentially more interesting application of this formalism is to
Drell-Yan rapidity distributions, which will be explored at very small
$x$ at the LHC, specifically by the LHCb
collaboration~\cite{McNulty:2009zz}, and which may play an important
role both in determining parton distributions
and testing resummed QCD predictions at small $x$. More in general,
our result provide a first generalization of small $x$ factorization
and resummation
beyond the inclusive LL$x$ level. It will be interesting to see
whether they can be generalized  not only to less inclusive
quantities, but also beyond the leading logarithmic approximation.

\vspace{0.5cm}

{\bf Acknowledgment} We thank R.~D.~Ball   for
several
discussions which are at the origin of  this paper and for a
critical reading of the manuscript.  
 F.C. acknowledges many useful
discussions with G.~Diana,   S.F. thanks V.~del Duca
for discussions during the early stage of this work,  and
S.M. thanks F.~Petriello for fruitful correspondence on the fixed-order
 results.
This work was supported in part by the European network HEPTOOLS under
contract MRTN-CT-2006-035505 and by an Italian PRIN-2008 grant.
The work of S.M. is supported by UK's STFC.
\clearpage

\appendix

\section*{Appendix}

\section{The $\msb$ subtraction procedure to all orders}\label{app_CFP}
Here we briefly sketch how to subtract collinear singularities arising
from kernel iteration within the $\msb$ scheme. 
Here we outline only the features which are relevant to understand how
Eqs.~\href{ngluon_msb},~\href{nkernel_msb} are obtained. For a general
treatment, including a discussion of
ultraviolet and infrared  singularities, 
as well as for a more detailed explanation,
we refer the reader to the original papers~\cite{EllisGeorgi, CFP}.

We start by writing a generic cross-section as the convolution 
product of a ``bare'' (i.e. unsubtracted)
coefficient function $\bar M$ and a ``bare'' PDF $\bar\Gamma$:
\be\label{CFP_cross}
\sigma = \bar M \otimes \bar\Gamma,
\ee
where $\otimes$ denotes the standard convolution product in $x$ space.
We assume to work in $d=4-2\epsilon$ dimensions in order to 
regularize $\bar M$ and $\bar \Gamma$. Following 
Sect.~\ref{inclusive}, we now write the coefficient function $\bar M$ as
the product of a hard part $H$ and the iteration of a kernel $K$:
\be\label{kernel_exp}
\bar M = H \otimes_{x,k_T} (1 + K + K\otimes_{x,k_T} K + ... ) \equiv 
M (1 + K + K^2 + K^3 + ... ) = H \frac{1}{1-K},
\ee
where now the convolution $\otimes_{x,k_T}$ stands both for normal 
convolution in $x$ space and for $k_T$ integration in the transverse space. 
This convolution product will be understood in the following formal 
manipulations. 

As a result of the $k_T$ integrations, the coefficient function 
Eq.~\href{kernel_exp} contains multiple collinear poles of the form
$1/\epsilon^j$ and thus it is meaningless in 4 dimensions. To perform the 
subtraction, we first introduce the ``pole-part'' projector $P$ which 
acting on a generic function $F$ selects just its $1/\epsilon^k$ poles, i.e.
\be\label{P_MS}
P F \equiv \sum_{k>0}\frac{1}{\epsilon^k}\lim_{\epsilon\rightarrow 0} 
\lp \epsilon^k F \rp.
\ee
We now use this projector to factorize the kernel $1/(1-K)$ in 
Eq.~\href{kernel_exp} into a finite part and a pure-pole part. To this
purpose, we first consider the kernel $(1-K)$ and perform the following formal
manipulations:
\be\label{geometric_K}
1-K = 1 - PK - (1-P)K = \left[
1-PK (1-(1-P)K)^{-1}
\right]
\left[1-(1-P)K\right],
\ee
which leads to (note the reverse order):
\be\label{sub_K}
\frac{1}{1-K} = 
\left[\frac{1}{1-(1-P)K}\right]
\left[\frac{1}{1-PK (1-(1-P)K)^{-1}}\right].
\ee

This is the desired factorized expression: the term in the first square
bracket is by construction free of collinear singularities, which have been 
moved to the second term. Using this factorized form, we can now rewrite 
the cross-section \href{CFP_cross} as
\be\label{coll_fact}
\sigma = \bar M \bar \Gamma = \left\{H \left[\frac{1}{1-(1-P)K}\right]\right\}
\left\{
\left[\frac{1}{1-PK (1-(1-P)K)^{-1}}\right]
\bar\Gamma \right\} = 
M \Gamma,
\ee 
with
\bea\label{sub_def}
M &\equiv& H \left[\frac{1}{1-(1-P)K}\right] \nonumber \\
\Gamma &\equiv& \left[\frac{1}{1-PK (1-(1-P)K)^{-1}}\right]
\bar \Gamma.
\eea
Now both $M$ and $\Gamma$ are finite, hence we can safely let $\epsilon=0$ 
and obtain a well-defined $d=4$ result. 
Note that this subtraction scheme is uniquely defined once 
the action of $P$ is specified. 
The subtraction of  the pure pole part (as in Eq.~\href{P_MS}) defines
the MS scheme. In the $\msb$ scheme one chooses to subtract also terms 
proportional to $-\gamma_E+\ln 4\pi$ coming from angular integrations. At leading logarithmic accuracy the action of $P$ is defined as:
\be\label{polepart_msb}
P_{\msb} ~ F = 
\sum_{k>0} \lp \lim_{\epsilon \rightarrow 0} \epsilon^k F \rp \times
\frac{1} {\epsilon^k} \exp\left[ k\, \epsilon\lp - \gamma_E + \ln 4\pi \rp
\right].
\ee

We now illustrate this subtraction procedure on the simple case of the
$t-$channel iteration of Altarelli-Parisi gluons. In this case we have
\bea
K &=&  \frac{(4\pi)^\epsilon}{\Gamma(1-\epsilon)}\int_0^{Q^2} \frac{d k_T^2}
{\lp k_T^2\rp^{1+\epsilon}} \as \lp \mu^2\rp ^\epsilon\gamma_0(N) = 
-\frac{1}{\epsilon} \frac{(4\pi)^{\epsilon}}{\Gamma(1-\epsilon)}
\lp \as \gamma_0(N)\rp \lp \frac{\mu^2}{Q^2}\rp^\epsilon,
\nonumber \\
K^2 &=& \frac{(4\pi)^\epsilon}{\Gamma(1-\epsilon)} 
\int_0^{Q^2} \frac{d k_{T,2}^2}{\lp k_{T,2}^2\rp^{1+\epsilon}}
\lp\as \lp \mu^2\rp ^\epsilon\gamma_0(N)\rp 
 \frac{(4\pi)^\epsilon}{\Gamma(1-\epsilon)} \times\nonumber\\
&\times& \int_0^{k_{T,2}^2} \frac{d k_{T,1}^2}{\lp k_{T,1}^2\rp^{1+\epsilon}}
\lp\as \lp \mu^2\rp ^\epsilon\gamma_0(N)\rp =
\frac{1}{2 \epsilon^2} 
\frac{(4\pi)^{2\epsilon}}{\Gamma^2(1-\epsilon)}
\lp \as \gamma_0(N)\rp^2 
\lp \frac{\mu^2}{Q^2}\rp^{2 \epsilon}\nonumber\\
\eea
and so on. Note that $K^2$ contains both single and double poles:
\bea
 \frac{K^2}{\as^2\gamma_0^2} &=&   \frac{1}{2\epsilon^2}
+\frac{\ln \lp\frac{\mu^2}{Q^2}\right)+\ln(4\pi)-\gamma_E}{\epsilon}
\nonumber + \lp \ln \lp \frac{\mu^2}{Q^2}\rp + \ln(4\pi) -2 \gamma_E \rp
\times\nonumber\\
&&\quad\times
\lp \ln \lp \frac{\mu^2}{Q^2}\rp + \ln(4\pi)\rp
-\frac{\pi ^2}{12}+\gamma_E^2+O\lp\epsilon\rp.
\eea

Careless subtraction of all poles and $\ln(4\pi)$, $\gamma_E$ terms 
would lead to the incorrect result
\be
K^2_{\text{sub}} = \as^2 \gamma_0^2(N) \ln^2 \frac{Q^2}{\mu^2}.
\ee
The correct $\msb$ result is instead obtained following 
the procedure explained above:
instead of subtracting poles from $K^2$ we should consider $(1-P)[K(1-P)K]$.
In this case we get
\bea
(1-P)K &=& \lp \as \gamma_0(N) \rp 
\lp -\frac{1}{\epsilon}\frac{(4\pi)^\epsilon}{\Gamma(1-\epsilon)}
\lp\frac{\mu^2}{Q^2}\rp^\epsilon + \frac{S_\epsilon}{\epsilon} \rp;
\nonumber \\
K(1-P) K &=& \lp \as \gamma_0(N)\rp^2 
\lp \frac{1}{2\epsilon^2}\frac{(4\pi)^{2\epsilon}}{\Gamma^2(1-\epsilon)}
\lp\frac{\mu^2}{Q^2}\rp^{2\epsilon} +\right.\\
&&-\left.
\frac{1}{\epsilon^2}\frac{(4\pi)^\epsilon}{\Gamma(1-\epsilon)}
\lp\frac{\mu^2}{Q^2}\rp^{\epsilon} S_\epsilon\rp;\nonumber \\
(1-P)\left[K(1-P)K\right] &=&
\lp \as \gamma_0(N)\rp^2 
\lp \frac{1}{2\epsilon^2}\frac{(4\pi)^{2\epsilon}}{\Gamma^2(1-\epsilon)}
\lp\frac{\mu^2}{Q^2}\rp^{2\epsilon} + \right. \nonumber\\
&&\left.-
\frac{1}{\epsilon^2}\frac{(4\pi)^\epsilon}{\Gamma(1-\epsilon)}
\lp\frac{\mu^2}{Q^2}\rp^{\epsilon} S_\epsilon+
\frac{1}{2\epsilon^2} S_\epsilon^2 \rp = \nonumber\\
&=&
\lp \as \gamma_0(N)\rp^2 
\frac{1}{2\epsilon^2}
\lp 
\frac{(4\pi)^\epsilon}{\Gamma(1-\epsilon)}\lp\frac{\mu^2}{Q^2}\rp^{\epsilon}
-S_\epsilon
\rp^2;
\eea
with $S_\epsilon \equiv \exp\left[\epsilon\lp \ln(4\pi) - \gamma_E\rp \right]$.
Note that this time
\be
(1-P)\left[K(1-P)K\right] = \frac{1}{2} 
\lp \as \gamma_0\rp^2
\ln^2 \frac{Q^2}{\mu^2} + O(\epsilon),
\ee
which is the correct $\msb$ result. 

This subtraction procedure can be easily
iterated. In particular, with $n$ kernels the result is
\bea
(1-P)K...(1-P)K &=& \lp \as \gamma_0 \rp^n
\frac{(-1)^n}{n!}\frac{1}{\epsilon^n}
\lp
\frac{(4\pi)^\epsilon}{\Gamma(1-\epsilon)}\lp\frac{\mu^2}{Q^2}\rp^{\epsilon}
-S_\epsilon
\rp^n = \nonumber\\
&=& \lp \as \gamma_0 \rp^n \frac{1}{n!}\ln^n\frac{Q^2}{\mu^2} + O(\epsilon).
\eea
Note that before performing the last $(1-P)$ subtraction, the $n$-kernel
result contains only a single $1/\epsilon^n$ pole. 

A particularly important feature of this subtraction method is its
iterative nature.  This is especially well suited for the 
purpose of this paper, because
we can perform all the universal kernel subtractions once for all, 
and consider separately the single collinear divergence coming from attaching 
the universal ladder to the process-dependent hard coefficient function, 
see Eqs.~\href{ngluon_msb},~\href{nkernel_msb}. 
Also, the 
$\msb$ subtraction factor $S_\epsilon$ exactly cancels all terms
coming from $(4\pi)^\epsilon/\Gamma(1-\epsilon)$:
the full $\msb$ result can be obtained by just
neglecting all $(4\pi)^\epsilon/\Gamma(1-\epsilon)$ terms and using 
$S_\epsilon = 1$. This simplification only relies on the fact 
that each emission is accompanied exactly by a 
$(4\pi)^\epsilon/\Gamma(1-\epsilon)$
term, which happens if there are no angular correlations. 
Since in all our derivations angular correlations are
subleading, in the main paper we have used this simplified version of the
subtraction procedure.

\section{Resummation of rapidity distributions in $\msb$}
\label{app_msbrap}
In this appendix we provide an explicit derivation of 
the resummation of the
rapidity distribution 
in the $\msb$ scheme. Because rapidity is determined by the
longitudinal momentum components, while collinear singularities appear
in transverse momentum integrations,
the main features of the $\msb$ derivation 
are the same of the derivation in Sec.~\ref{rapidity}.

For simplicity we consider emissions from just one leg. Extension to 
the full case is straightforward.
We start by writing the $n-$kernel fully differential cross-section 
in $d=4-2\epsilon$ dimensions, see Eq.~\href{nkernel_kt}
\begin{footnote}{Here again we omit all $(4\pi)^\epsilon/\Gamma(1-\epsilon)$
terms in view of  the $\msb$ subtraction, as discussed in 
Appendix~\ref{app_CFP}.}\end{footnote}:
\bea\label{diff_msb}
&&\d\bar\sigma^n
\lp x, \frac{\mu^2}{Q^2}, \as, z_i, \xi_i; \epsilon,\rp=\nonumber\\ &&
P\lp z_n, \lp\frac{\mu^2}{Q^2}\rp^\epsilon, \as; \epsilon \rp
\frac{dz_n}{z_n}\frac{d\xi_n}{\xi_n^{1+\epsilon}}
\times
C\lp x, z_i, \xi_i, \as; \epsilon\rp \times \nonumber\\ 
&&\times P\lp z_{n-1}, \lp\frac{\mu^2}{Q^2}\rp^\epsilon, \as; \epsilon \rp
\frac{dz_{n-1}}{z_{n-1}}\frac{d\xi_{n-1}}{\xi_{n-1}^{1+\epsilon}}
\times ... \times 
P\lp z_1 , \lp\frac{\mu^2}{Q^2}\rp^\epsilon, \as; \epsilon \rp
\frac{dz_{1}}{z_{1}}\frac{d\xi_{1}}{\xi_{1}^{1+\epsilon}},\nonumber\\
\eea
where as in Eq.~\href{1leg_fac} the splitting function $P$ is the inverse 
Mellin transform of the
anomalous dimension $\gamma$ in Eq.~\href{nkernel_kt}. Note that in 
Eq.~\href{diff_msb} we have written $C$ as the most generic (2PI) function
of all the relevant variables. Quasi-collinear 
kinematics determines the dependence of $C$ on its arguments.
Because  rapidity
distributions do not factorize in $N-$Mellin space, for the time being
we write everything in $x-$space. 

The $d-$dimensional rapidity 
distribution is immediately obtained from the fully differential
cross-section Eq.~(\ref{diff_msb}):
\bea\label{dy_kt_msb}
&&\frac{\d \bar \sigma^n}{\d y}
\lp x,\frac{\mu^2}{Q^2},y, \as; \epsilon\rp = \nonumber\\
&&\quad=\int_x^1\frac{dz_n}{z_n}
P\lp z_n, \lp\frac{\mu^2}{Q^2}\rp^\epsilon, \as; \epsilon \rp\times ...
\times
\int_{\frac{x}{z_2 ... z_n}}^1\frac{dz_1}{z_1}
P\lp z_1, \lp\frac{\mu^2}{Q^2}\rp^\epsilon, \as; \epsilon \rp\times
\nonumber\\
&&\qquad\times
\int_0^\infty \frac{d\xi_n}{\xi_n^{1+\epsilon}}
C_y\lp x, z_i, \xi_i, y, \as; \epsilon\rp
\int_0^{\xi_n}\frac{d\xi_{n-1}}{\xi_{n-1}^{1+\epsilon}}\times...\times
\int_0^{\xi_2}\frac{d\xi_{1}}{\xi_{1}^{1+\epsilon}}, \nonumber\\
\eea
where the rapidity-dependent coefficient function is defined as
\bea\label{cy_msb}
&&C_y\lp x, z_i, \xi_i, y, \as; \epsilon\rp
\equiv \frac{Q^2}{2 s z_1...z_n}\int_0^{2\pi}\frac{d\theta}{2\pi}
\int d\Pi_{\mathcal F} 
\overline{
\left|\mathcal M\lp \mathcal V(n)+ g^*(p_L)\rightarrow 
\mathcal F\rp
\right|^2}\times\nonumber\\
&&\quad \times \delta_4(P_I-P_f)\times
\delta\lp y - \frac{1}{2}\ln \frac{E_{\mathcal S}+p_{\mathcal S z}}
{E_{\mathcal S}-p_{\mathcal S z}}\rp,
\eea
in terms of 
the spin and color averaged
squared amplitude $\overline{\left|\mathcal M\right|^2}$ 
for the off-shell process $
\mathcal V(n)+g^*(p_L)\rightarrow \mathcal F$ (see Fig.~\ref{hard_pic}),
and the polarization sum for the off-shell gluon is performed through the 
projector $\mathcal P$ Eq.~\href{proj}. We now use the fact that 
we are in  quasi-collinear kinematics (i.e. $\xi_i\ll \xi_{i+1}$) and
in the small $x$ regime Eq.~\href{sxkin}. This implies first, that $C_y$
is sensitive only to the largest transverse momenta $\xi_n$ 
and second, that up to terms which vanish when 
$k_T^2\ll s$ the effect of the ladder insertion to $C_y$ is just a longitudinal
boost. We have thus
\bea\label{cy_boost_msb}
&&C_y(x,z_i,\xi_i,y,\as;\epsilon) = \nonumber\\
&&\quad = 
C_y\lp\frac{x}{z_1...z_n}, \xi_n, 
y-\frac 1 2 \ln z_1 - ... - \frac{1}{2} \ln z_n, \as;
\epsilon\rp + O\lp \frac{k_T^2}{s}\rp.
\eea
Since upon integration $O(k_T^2/s)$ terms give rise to contributions suppressed
by powers of $x$, we can safely discard them. 

Thanks to 
 quasi-collinear kinematics we can thus write Eq.~\href{dy_kt_msb} as
\bea\label{dy_kt_msb_coll}
&&\frac{\d \bar \sigma^n}{\d y}
\lp x,\frac{\mu^2}{Q^2},y, \as; \epsilon\rp = \nonumber\\
&&=\int_x^1\frac{dz_n}{z_n}
P\lp z_n, \lp\frac{\mu^2}{Q^2}\rp^\epsilon, \as; \epsilon \rp\times ...
\times
\int_{\frac{x}{z_2 ... z_n}}^1\frac{dz_1}{z_1}
P\lp z_1, \lp\frac{\mu^2}{Q^2}\rp^\epsilon, \as; \epsilon \rp\times
\nonumber\\
&&\times
\int_0^\infty \frac{d\xi_n}{\xi_n^{1+\epsilon}}
C_y\lp \frac{x}{z_1...z_n}, \xi_n, y-\frac{1}{2} \ln z_1-...-
\frac{1}{2}\ln z_n, \as; \epsilon\rp\times\nonumber\\
&&\times
\int_0^{\xi_n}\frac{d\xi_{n-1}}{\xi_{n-1}^{1+\epsilon}}\times...\times
\int_0^{\xi_2}\frac{d\xi_{1}}{\xi_{1}^{1+\epsilon}} + O(x). 
\eea
From now on we will systematically omit all $O(x)$ terms.

Convolution products in Eq.~\href{dy_kt_msb_coll}  turn into ordinary
ones by Fourier-Mellin transformation:
\bea
&& \frac{\d \bar \sigma^n}{\d y} \lp N, \frac{\mu^2}{Q^2}, b, \as; \epsilon\rp
= \left[\gamma \lp N - \frac{ib} 2, \lp \frac{\mu^2}{Q^2} \rp ^\epsilon, \as;
\epsilon\rp \right]\times\nonumber\\
&&\quad\times \int_0^\infty \frac{d\xi_n}{\xi_n^{1+\epsilon}} 
C_y\lp N, \xi_n, b, \as; \epsilon\rp
\times\nonumber 
\int_0^{\xi_n} 
\left[\gamma \lp N - \frac{ib} 2, \lp \frac{\mu^2}{Q^2} \rp ^\epsilon, \as;
\epsilon\rp \right]\frac{d\xi_{n-1}}{\xi_{n-1}^{1+\epsilon}} \times\nonumber
\\
&&\quad\times ... \times
\int_0^{\xi_2}\left[\gamma \lp N - \frac{ib} 2, \lp \frac{\mu^2}{Q^2} 
\rp ^\epsilon, \as; \epsilon\rp \right]\frac{d\xi_{1}}{\xi_{1}^{1+\epsilon}}.  
\eea
This expression is identical to Eq.~\href{nkernel_kt}, but with
the anomalous dimension $\gamma$ is evaluated at $N-i b/2$. This
does not interfere with collinear singularities, hence 
the $\msb$ subtraction is performed
as in the inclusive case, with the result (see Eq.~\href{kt_resfull})
\bea\label{res_msb_sub}
&&\frac{\d \sigma}{\d y} = 
\sum_{n=1}^\infty \frac{\d \bar\sigma^n}{\d y} = 
\gamma\lp N - \frac{ib} 2, \lp \frac{\mu^2}{Q^2}\rp^\epsilon, \as; \epsilon\rp
\int_0^\infty \frac{d \xi}{\xi^{1+\epsilon}}
C_y\lp N, \xi, b, \as; \epsilon\rp
\times \nonumber \\&&
\times \exp\left[
\frac{1}{\epsilon} \sum_k \frac{\tilde\gamma_k\lp N- \frac{ib} 2, \as; 0\rp}{k} 
\lp 1 - \lp \frac{\mu^2}{Q^2\xi}\rp^{k \epsilon} 
\frac{\tilde\gamma_k(N-i b/2,\as;\epsilon)}{\tilde\gamma_k(N-i b/2,\as;0)}\rp 
\right],\nonumber\\
\eea
where $\tilde \gamma_k$ was defined in Eq.~\href{andimexp}.
As in the inclusive case, Eq.~\href{res_msb_sub} has a finite $\epsilon\to0$ 
limit:
\bea\label{to0_msb_R}
\lim_{\epsilon\rightarrow 0} & &\frac{1}{\xi^{1+\epsilon}}
\exp\left[
\frac{1}{\epsilon} \sum_i \frac{\asb^i}{i} \gamma_i\lp N- \frac{ib} 2, 0\rp
\lp 1 - \lp \frac{\mu^2}{Q^2\xi}\rp^{i \epsilon} 
\frac{\gamma_i(N-i b/2,\epsilon)}{\gamma_i(N-i b/2,0)}\rp 
\right]= \nonumber \\
&=&\xi^{\gamma\lp N - \frac{ib} 2, \as\rp -1}
\exp\left[\gamma\lp N - \frac{ib} 2, \as \rp \ln \frac{Q^2}{\mu^2}\right]
e^{-\sum_i \frac{\as^i \dot \gamma_i(N-i b/2)}{i}}=\nonumber \\
&=&\xi^{\gamma\lp N - \frac{ib} 2, \as\rp -1}
\exp\left[\gamma\lp N - \frac{ib} 2, \as \rp \ln \frac{Q^2}{\mu^2}\right]
\mathcal R \lp N - \frac{ib} 2, \as \rp.
\eea

Using Eq.~(\ref{to0_msb_R}) the resummed result immediately follows 
(with $\mu^2=Q^2$, thus omitting the explicit dependence on $\mu^2/Q^2$):
\bea\label{dy_res_msb_r}
\frac{\d \sigma}{\d y}(N,\as(Q^2), b) 
= \gamma \lp N - \frac{ib} 2, \as(Q^2)\rp
\times\nonumber\\ \times
\int_0^\infty d\xi \xi^{\gamma\lp N-i b/2, \as(Q^2)\rp-1}
C\lp N,\xi, \as(Q^2), b\rp
\mathcal R\lp N - \frac{ib} 2, \as(Q^2)\rp,
\eea
which is the same of Eq.\href{l_iter0} 
but this time with a shifted $\mathcal R$ factor. Note that at LL$x$ we
also have
\be\label{andim_sx}
\gamma\lp N-\frac{ib}{2}, \as \rp = \gamma\lp \frac{\as}{N-i b/2}\rp;
\qquad \mathcal R\lp N-\frac{ib}{2}, \as \rp 
= \mathcal R\lp \frac{\as}{N-i b/2}\rp.
\ee

We finally show how  to deal with
a scheme change in rapidity distribution, thereby in particular
explaining  the shift in the argument 
of $\mathcal R$ in Eq.~\href{dy_res_msb_r}. 
To this purpose, assume a scheme change from $\msb$ to some ``small
$x$'' scheme is performed by dividing the
parton distribution by  a LL$x$ function ${\mathcal N}$:
\be\label{scheme_def}
M_{\rm{small}~x} \equiv M_{\msb} \mathcal N;\qquad 
\Gamma_{\rm{small}~x} \equiv \frac{1}{\mathcal N}\Gamma_{\msb}.
\ee
Using the $\msb$ kernel expansion Eq.~\href{sub_def} we can rewrite
$M_{\rm{small}~x}$ as
\be\label{hexp_scheme}
M_{\rm{small}~x} = H \frac{1}{1-(1-P) K} \mathcal N = 
H \left[1 + (1-P) K + (1-P)\left[K(1-P)K\right]+...\right] \mathcal N.
\ee
where (see Appendix~\ref{app_CFP}) all products must be understood 
as convolution in $x$ and $k_T$ spaces.
It follows that the effect of the scheme change is the same as that of
a kernel insertion on the first (outer) rung of the ladder:
\bea\label{dy_gnorm}
&&\frac{\d \bar\sigma^1}{\d y} 
\lp x, \frac{\mu^2}{Q^2}, y, \as;\epsilon\rp
= \int_x^1 \frac{d z}{z} \mathcal N\lp z, \as; \epsilon\rp \times\nonumber\\
&&\quad\times
\int_{x/z} \frac{d z_1}{z_1} P \lp z_1, \as, \lp \frac{\mu^2}{Q^2}\rp^\epsilon;
\epsilon\rp \int_0^\infty\frac{d\xi}{\xi^{1+\epsilon}} 
C\lp \frac{x}{z z_1},\xi, y-\frac{1}{2}\ln z - \frac{1}{2}\ln z_1, \as;
\epsilon\rp.\nonumber\\
\eea
In  Fourier-Mellin space the convolutions in Eq.~\href{dy_gnorm} 
turn into ordinary products 
\bea
&&\frac{\d \bar\sigma^1}{\d y} \lp N, \frac{\mu^2}{Q^2}, b, \as;\epsilon\rp = 
\mathcal N \lp N - \frac{ib}{2}, \as; 
\epsilon\rp\times
\nonumber\\
&&\quad \times \gamma\lp N- \frac{ib}{2},\as, \lp \frac{\mu^2}{Q^2}\rp^\epsilon; 
\epsilon\rp \int_0^\infty \frac{d\xi}{\xi^{1+\epsilon}}
C_y\lp N -  \frac{ib}{2}, b, \as;\epsilon
\rp,
\eea
so the argument of $\mathcal N$ ends up being also  shifted when used
for computing rapidity distributions. Clearly this holds also if we
consider the full ladder and not only a single kernel insertion. 

Collecting the results of this section, we can write the full $\msb$ 
rapidity distribution resummation formula as
\bea\label{res_full_msb}
\frac{\d \sigma}{\d y}\lp N,\frac{\mu^2}{Q^2},b, \as(\mu^2)\rp
= \left[\gamma \lp\frac{\as(\mu^2)}{ N - \frac{ib} 2}\rp
\exp\left[\gamma\lp\frac{\as(\mu^2)}{ N - \frac{ib}{2}}\rp
\ln \frac{Q^2}{\mu^2}\right]
\times\right.\nonumber\\ \times\left.
\int_0^\infty d\xi \xi^{\gamma\lp \frac{\as(\mu^2)}{N-i b/2}\rp-1}
C\lp N,\xi, b, \as(\mu^2)\rp\right]
R\lp \frac{\as(\mu^2)}{N - \frac{ib} 2}\rp.
\eea
The extension of Eq.~\href{res_full_msb} to emissions from both legs along
the lines of Sec.~\ref{rapidity} is straightforward
\bea\label{res_full_msb_2legs}
&&\frac{\d \sigma}{\d y}\lp N,\frac{\mu^2}{Q^2},b, \as(\mu^2)\rp = 
\gamma \lp\frac{\as(\mu^2)}{ N - \frac{ib} 2}\rp
\exp\left[\gamma\lp\frac{\as(\mu^2)}{ N -  \frac{ib}{2}}\rp
\ln \frac{Q^2}{\mu^2}\right]
\times\nonumber\\
&&\quad\times\gamma \lp\frac{\as(\mu^2)}{ N + \frac{ib} 2}\rp
\exp\left[\gamma\lp\frac{\as(\mu^2)}{ N + \frac{ib}{2}}\rp
\ln \frac{Q^2}{\mu^2}\right]
\times R\lp \frac{\as(\mu^2)}{N - \frac{ib} 2}\rp
R\lp \frac{\as(\mu^2)}{N + \frac{ib} 2}\rp\times
\nonumber \\
&&\quad\times
\int_0^\infty d\xi \xi^{\gamma\lp \frac{\as(\mu^2)}{N-i b/2}\rp-1}
\int_0^\infty d\bar\xi \bar\xi^{\gamma\lp \frac{\as(\mu^2)}{N+i b/2}\rp-1}
C\lp N,\xi,\bar\xi, b, \as(\mu^2)\rp.
\eea

\end{document}